\documentclass[reprint,twocolumn,superscriptaddress,showpacs,showkeys,amsmath,amssymb,floatfix,aps,prb]{revtex4-1}

\usepackage{graphicx}
\usepackage{dcolumn}
\usepackage{bm}
\usepackage{amsmath}
\usepackage{placeins}
\usepackage{color}
\usepackage{hyperref}
\usepackage{multirow}
\usepackage{pdfpages}
\makeatletter
\AtBeginDocument{\let\LS@rot\@undefined}
\makeatother

\begin{document}
\title{\texorpdfstring{Proximity exchange effects in MoSe$_2$
and WSe$_2$ heterostructures with  CrI$_3$: twist angle, layer, and
gate dependence}{}}

\author{Klaus Zollner}
\email{klaus.zollner@physik.uni-regensburg.de}
\affiliation{Institute for Theoretical Physics, University of Regensburg, 93040 Regensburg, Germany}

\author{Paulo E. Faria~Junior}
\affiliation{Institute for Theoretical Physics, University of Regensburg, 93040 Regensburg, Germany}

\author{Jaroslav Fabian}
\affiliation{Institute for Theoretical Physics, University of Regensburg, 93040 Regensburg, Germany}

\date{\today}

\begin{abstract}
Proximity effects in two-dimensional (2D) van der Waals heterostructures offer controllable ways to tailor the electronic band structure of adjacent materials. Proximity exchange in particular is important for making materials magnetic without 
hosting magnetic ions. Such {\it synthetic} magnets
could be used for studying magnetotransport in high-mobility 2D materials, or magneto-optics in 
highly absorptive nominally nonmagnetic semiconductors. Using first-principles calculations, we show that the proximity exchange
in monolayer MoSe$_2$ and WSe$_2$ due to ferromagnetic monolayer CrI$_3$ can be tuned (even qualitatively) by twisting and gating. Remarkably,
the proximity exchange remains the same when 
using antiferromagnetic CrI$_3$ bilayer, paving the way for optical and electrical detection of layered antiferromagnets. Interestingly, the proximity exchange is {\it opposite} to the exchange of the adjacent antiferromagnetic layer. Finally, we show that the proximity exchange is confined to the layer adjacent to CrI$_3$, and that adding a separating hBN barrier drastically reduces the proximity effect. 
We complement our {\it ab initio} results with tight-binding modeling and solve the Bethe-Salpeter equation to provide experimentally verifiable optical signatures (in the exciton spectra) of the proximity exchange effects.  
\end{abstract}

\maketitle

\section{Introduction}
Two-dimensional (2D) materials and their hybrids offer unforeseen
opportunities, but also challenges, to 
electronics, spintronics, optics and magnetism 
\cite{Fiori2014:NN,Ferrari2015:NS,Choi2017:MT, Schwierz2015:NS, Zutic2004:RMP}. Graphene, the prototypical 2D crystal, 
has excellent charge and spin transport properties \cite{Gurram2017:2DM, Neto2009:RMP, Sarma2011:RMP, Han2014:NN}, 
but lacks an orbital band gap needed for digital transistor applications. Fortunately, we have now available 
a large class of air-stable 2D semiconductors---transition-metal dichalcogenides (TMDCs)---which have a band gap in the optical range 
\cite{Kormanyos2014:2DM, Liu2015:CSR, Tonndorf2013:OE,Tongay2012:NL,Eda2011:NL}, and form a favorite 
platform for optical experiments including optical spin injection
due to helicity-selective optical exitations\cite{Xiao2012:PRL}:
electrons in opposite valleys, but at the same energy, feel opposite spin-orbit fields, pointing out of the plane. This effect is called
valley Zeeman coupling; unlike true Zeeman field, the valley coupling preserves time reversal symmetry, since it stems from spin-orbit coupling (SOC).

In fact, one can create a {\it proximity structure} from graphene and a TMDC, initially
proposed by DFT calculations \cite{Gmitra2015:PRB} and confirmed
experimentally \cite{Luo2017:NL, Avsar2017:ACS} to facilitate transfer
of the optically generated spin in TMDC into graphene. This is an example
of a proximity effect \cite{Zutic2018:MT} in van der Waals heterostructures. Proximity effects provide fascinating opportunities for {\it band-structure engineering}. Experiments and theory show that graphene can borrow different properties from a variety of substrates, be it SOC or magnetism \cite{Gmitra2015:PRB, Zollner2016:PRB, Qiao2014:PRL, Zhang2018:PRB, Averyanov2018:ACS, Khokhriakov2018:SCA, Hallal2017:2DM, Cardoso2018:PRL}.

For {\it all-2D spintronics}, it is desirable to integrate 2D materials such as graphene and TMDCs with 2D magnets. Experimentalists have demonstrated magnetic order in 2D layered crystals, such as MnSe$_2$ \cite{OHara2018:NL, Kan2014:PCCP}, CrGeTe$_3$ 
\cite{Li2014:JMCC, Carteaux1995:JP, Gong2017:Nat, Siberchicot1996:JPC, Lin2017:PRB, Wang2018:NN}, and CrI$_3$ 
\cite{Liu2016:PCCP, Zhang2015:JMCC, McGuire2015:CoM, Webster2018:PCCP, Huang2017:Nat, Jiang2018:NL, Soriano2018:arxiv,Huang2018:NN, Jiang2018:NN, Wu2019:NC}, which are well suited for nanoelectronic devices \cite{Burch2018:Nat}. Unlike thin films of conventional ferromagnets, which have magnetization typically in the plane (of the film), the 2D layered ferromagnets have magnetization pointing out of the plane, making them Ising-like. 

This out-of-plane exchange interaction 
is a time-reversal breaking analog of the valley Zeeman splitting in TMDCs. The interplay of the two couplings, 
exchange and valley Zeeman, motivates explorations of 
stacked TMDCs and 2D ferromagnets. Certainly, one can introduce 
Zeeman coupling by applying an external magnetic field pointing out of the plane, but such fields produce modest valley splittings, about 0.1 - 0.2 meV per tesla\cite{Srivastava2015:NP, Aivazian2015:NP, Li2014i:PRL, MacNeill2015:PRL}. Proximity exchange fields 
can induce much stronger effects, perhaps up to hundreds of meVs, without significantly altering the band structure of TMDCs. 
\cite{Vitale2018:S, Langer2018:Nat,Schaibley2016:NRM, Ye2016:NN, Zhong2017:SA, Ji2018:PCCP, Li2018:PCCP, Qi2015:PRB, Zhang2016:AM, Xu2018:PRB, Zhao2017:NN, Ye2016:NN, Zhong2017:SA, Seyler2018:NL, Peng2017:ACS}. Conventional ferromagnetic substrates, such as EuO or MnO, were predicted to give 200 - 300 meV \cite{Qi2015:PRB, Zhang2016:AM, Xu2018:PRB}; 
experiments on EuS find only 2.5 meV  \cite{Zhao2017:NN}, presumably due to uneven interfaces.

There already are experiments demonstrating proximity exchange
in TMDCs. Recent measurements in TMDC/CrI$_3$ 
heterostructures \cite{Seyler2018:NL, Zhong2017:SA} show a few meV of proximity exchange. 
CrI$_3$ is especially interesting, because the monolayer is a ferromagnet (FM) \cite{Jiang2018:arxiv, Huang2017:Nat}, 
while bilayer CrI$_3$ shows antiferromagnetic (AFM) coupling \cite{Huang2017:Nat, Jiang2018:NM, Song2018:SC}, 
in contradiction to the existing theory \cite{Jiang2018:arxiv, Soriano2018:arxiv, Sivadas2018:NL} which predicts a FM state for the low temperature phase.
Remarkably, the magnetization of the CrI$_3$ can 
be tuned optically \cite{Seyler2018:NL}, thereby influencing proximity 
exchange and the photoluminescence (PL) spectrum of the TMDC. 
In addition, the magnetism in few layer CrI$_3$ can be controlled 
by gating and external magnetic fields \cite{Jiang2018:NN, Huang2018:NN}, 
opening a new path for gate controlled devices, such as 
spin-filter tunnel junctions \cite{Song2018:SC, Klein2018:SC, Ghazaryan2018:NE}, 
and AFM spintronics \cite{Baltz2018:RMP, Jungwirth2016:NN}.

Here, we provide a systematic theoretical analysis
of the proximity exchange coupling in
TMDC/CrI$_3$ heterostructures (with MoSe$_2$ and
WSe$_2$ as TMDC) from first-principles. First, we confirm that the magnetic insulator substrate couples weakly to the TMDC by van der Waals forces, preserving the characteristic electronic band structure of the TMDC. The proximity exchange coupling splits the conduction (CB) and valence band (VB) of the TMDC by roughly 1--5 meV, and combined 
with the intrinsic (valley Zeeman) SOC of the TMDC lifts the valley degeneracy. We 
introduce a minimal model Hamiltonian to describe the proximity effects in TMDC due to CrI$_3$,  extracting realistic proximity exchange parameters which should be useful for modeling transport and optics. 

Next, and this is the main result of the paper, we find wide tunability of proximity exchange effects with respect of twisting and gating, and the absence of effects coming from additional layers
(both TMDC and CrI$_3$). 
In particular, we find that {\it proximity exchange splittings depend on the twist angle} between the TMDC and the CrI$_3$. We investigated 0 and 30 degrees structures, and observed that not only the magnitudes of the exchange differ, but, remarkably, the direction of the exchange field for holes changes sign. The exchange parameters can be tuned by a few meVs by gating, using
accessible electric fields of a few V/nm. 

It is rather fascinating that adding another layer
of CrI$_3$ does not affect the proximity exchange
in TMDCs (we used MoSe$_2$), given that the two magnetic layers are
antiferromagnetically coupled and they have zero net magnetic moment. The proximity effect can then be used to detect, optically or electrically, the magnetic moment of the adjacent (to the TMDC) CrI$_3$ layer {\it even in the antiferromagnetic state}.
We also explicitly prove the short-range nature of the proximity effects by investigating
bilayer-MoSe$_2$/CrI$_3$ and MoSe$_2$/hBN/CrI$_3$.
In the former, the proximity affects the adjacent TMDC layer, while in the latter the insulating hBN layer
drastically reduces the proximity exchange. This message is important since experimentally it may be desirable to cover CrI$_3$ with hBN first, to
improve its stability under ambient conditions. For proximity effects, using hBN barriers would be detrimental. 

Finally, we give specific predictions for 
optical signatures of the proximity exchange
effects, by calculating the excitonic absorption
spectra employing the Bethe-Salpeter equation. 
The twist-angle and gate-bias dependence of the proximity exchange is mapped into the valley splitting of the first intralayer exciton peak. 
This is a valuable and experimentally testable fingerprint of our results.

\section{Band structure, geometry, and twist effects}
\begin{figure*}[!htb]
 \includegraphics[width=.95\textwidth]{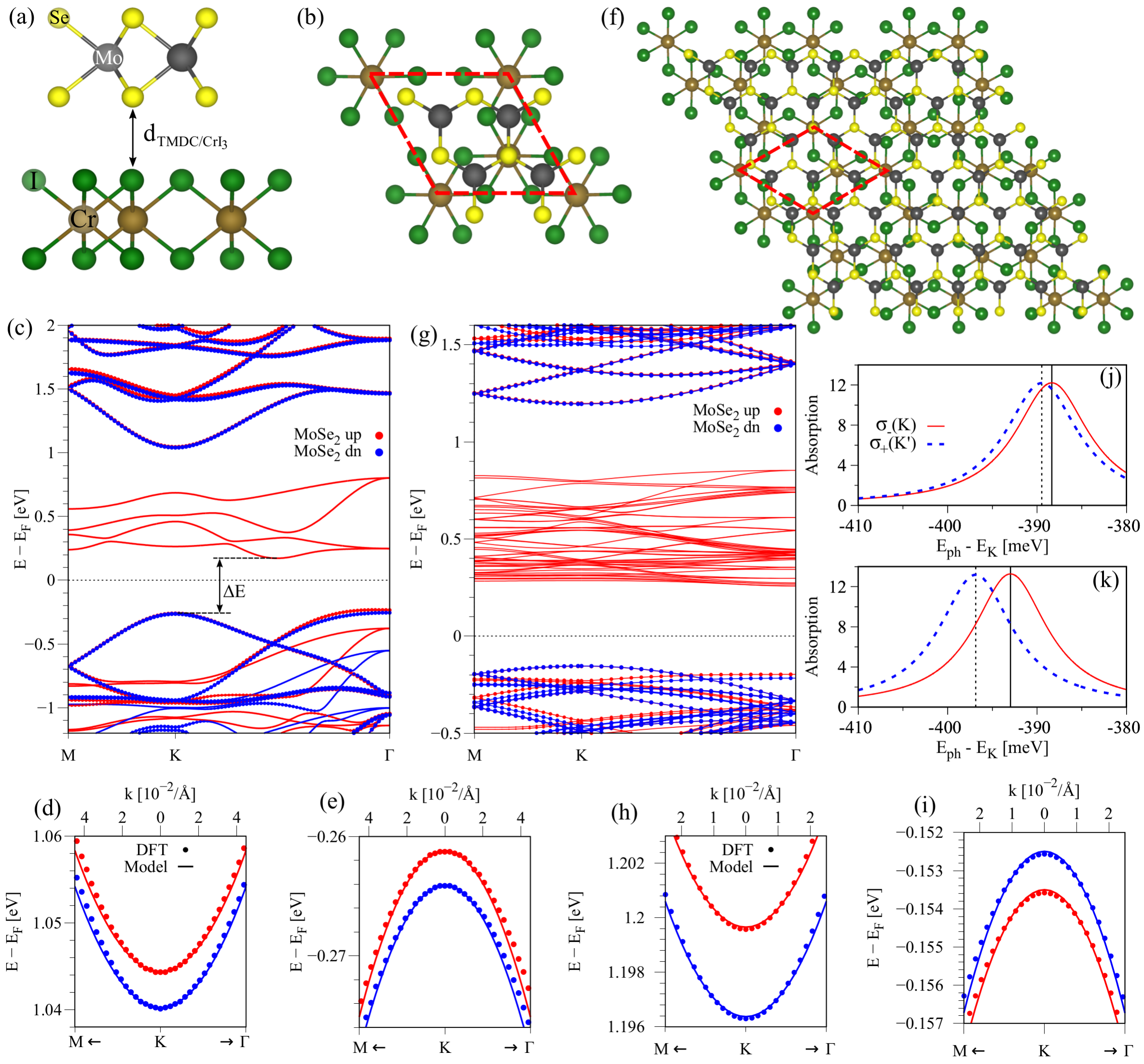}
 \caption{(Color online) Calculated band structures without SOC, geometries, and absorption spectra of MoSe$_2$/CrI$_3$. (a,b) Side and top view of the small supercell geometry ($0^{\circ}$ twist angle)
 with labels for the different atoms and the definition of the interlayer distance d$_{\textrm{TMDC/CrI$_3$}}$. One unit cell of CrI$_3$ is highlighted, by the red dashed line, in the top view. (c) Band structure along high symmetry lines, with the energy gap of the heterostructure $\Delta$E. 
 The bands corresponding to MoSe$_2$ are
 emphasized by red (spin up) and blue (spin down) spheres. 
 (d,e) Zoom to the CB and VB edge corresponding to MoSe$_2$. 
 Symbols are DFT data and solid lines are the fitted model Hamiltonian. 
 (f-i) The top view, calculated band structure, and the zoom to the low energy bands for the larger supercell geometry ($30^{\circ}$ twist angle). (j,k) Calculated first absorption peak of intralayer excitons for small (b) and large (f) geometry (including SOC) with the vertical solid (dashed) arrows indicating the peak position for the absorption at K (K') point.
 }\label{Fig:Figure1}
\end{figure*}

To study proximity exchange effects, we set up a common unit cell for the TMDC/CrI$_3$ heterostructures \footnotemark[1]. 
We consider a $2\times 2$ supercell for the TMDCs (MoSe$_2$ and WSe$_2$) above a $1 \times 1$ cell of CrI$_3$, as well as a larger $7\times 7$ supercell of the TMDC on top of a ($2\sqrt{3}\times2\sqrt{3}$)R$30^{\circ}$ supercell of CrI$_3$.
In Figs. \ref{Fig:Figure1}(a,b), we show the geometry of the MoSe$_2$/CrI$_3$ heterostructure, as a typical structure of the small supercell with $0^{\circ}$ twist angle. 
In Fig. \ref{Fig:Figure1}(f), we show the top view of the larger supercell which is 
twisted by $30^{\circ}$ relative to the underlying CrI$_3$.

In Figs. \ref{Fig:Figure1}(c-e) we show the calculated band structure 
with a fit to our model Hamiltonian \footnotemark[1] for the small MoSe$_2$/CrI$_3$ supercell structure without SOC (to extract the exchange coupling). 
We find that the bands of the TMDC are nicely preserved but are marked with a proximity exchange.
The spin up CBs, originating from the CrI$_3$, are located within the band gap of the TMDC, see Fig. \ref{Fig:Figure1}(c).
In experiments it was already shown that the spin polarized in-gap states from the CrI$_3$ 
quench the PL spectrum for one light helicity only \cite{Seyler2018:NL,Zhong2017:SA}, 
due to additional nonradiative relaxation processes. 
The energy gap of the full heterostructure is $\Delta E \approx 400$~meV, as defined in Fig. \ref{Fig:Figure1}(c). 
The band edges of the TMDC can be almost perfectly described by our model Hamiltonian, 
as shown in Figs. \ref{Fig:Figure1}(d,e).
Due to proximity exchange, the bands are spin split by about $5$~meV. 

In Figs. \ref{Fig:Figure1}(g-i) we present the calculated band structure 
with a fit to our model Hamiltonian for the larger MoSe$_2$/CrI$_3$ supercell structure with $30^{\circ}$ twist angle without SOC. 
As there are much more atoms in the supercell, more in-gap states from the CrI$_3$ are located within the band gap of the TMDC. The proximity exchange is clearly
visible. In comparison to the smaller, $0^{\circ}$ supercell,  
the VB splitting is {\it opposite 
in sign}. Unfortunately, studying other, especially small twist angles is beyond our DFT approach. But the different direction of the exchange field in VB seen for
$0^{\circ}$ and $30^{\circ}$ structures shows that twisting can be an effective
tool to modify the proximity effect. Similar effect was predicted in SOC proximity effect, by placing graphene on a topological insulator. Two different twist angles produced qualitatively different spin-orbit fields in graphene \cite{Song2018:NL}.

To investigate the optical signatures of the proximity exchange 
due to the twist angle, 
we calculate the absorption spectra of the intralayer excitons, i.e., electron-hole 
pairs created directly at the TMDC layer that are experimentally accessible in 
PL spectra\cite{Zhong2017:SA,Seyler2018:NL}. For these calculations we apply the 
effective Bethe-Salpeter equation\cite{RohlfingPRB:2000,Scharf2017:PRL,Tedeschi2019:PRB,FariaJunior2019phosphorene} using the model Hamiltonian 
fitted to the first-principles band structure, see Figs.~\ref{Fig:Figure1}(d,e,h,i). 
For these calculations we also consider the effects of SOC by combining the parameters 
summarized in Tab.~\ref{Tab:fit_TMDC_CrI3} (see Supplemental Material \footnotemark[1] for details of the 
excitonic calculations and the model Hamiltonian). Focusing on the lowest energy excitonic 
levels, we show in Figs.~\ref{Fig:Figure1}(j,k) the first absorption peak at K ($\sigma_-$ polarization) 
and K' ($\sigma_+$ polarization) valleys for $0^{\circ}$ and $30^{\circ}$ twist angles. 
We find that the valley splitting (the energy separation between the two absorption peaks) shows a {\it substantial 
3-fold increase} by changing the twist angle, from $\sim$1.13 meV for 
$0^{\circ}$ to $\sim$3.89 meV for $30^{\circ}$.
Interestingly, these energy splittings 
calculated within the single-particle picture ($\sim$1.25 meV for $0^{\circ}$ and 
$\sim$4.29 meV for $30^{\circ}$) are in reasonable agreement with the excitonic calculations. We also performed a similar investigation for WSe$_2$/CrI$_3$ heterostructures and found an excitonic (single-particle) valley splitting of $\sim$1.43 (1.61) meV for $0^{\circ}$ and $\sim$6.35 (7.09) meV for $30^{\circ}$, thus providing a dramatic $\sim$4.4-fold increase due to twist angle. 
In recent experiments by Zhong et al.\cite{Zhong2017:SA}, the measured valley splitting in WSe$_2$/CrI$_3$ is $\sim$3.5 meV (equivalent to $\sim$13 T external magnetic field in bare WSe$_2$), 
and thus it is reasonable to expect that the structure (or part of it, depending on the quality) was twisted. Undoubtedly, the WSe$_2$/CrI$_3$ heterostructure demands further experimental investigations of proximity exchange, especially with respect to different values of the twist angle. A similar twist angle dependence of proximity SOC has been  
reported for graphene/TMDC heterostructures \cite{Li2019:PRB}.

\begin{figure}[htb]
 \includegraphics[width=.99\columnwidth]{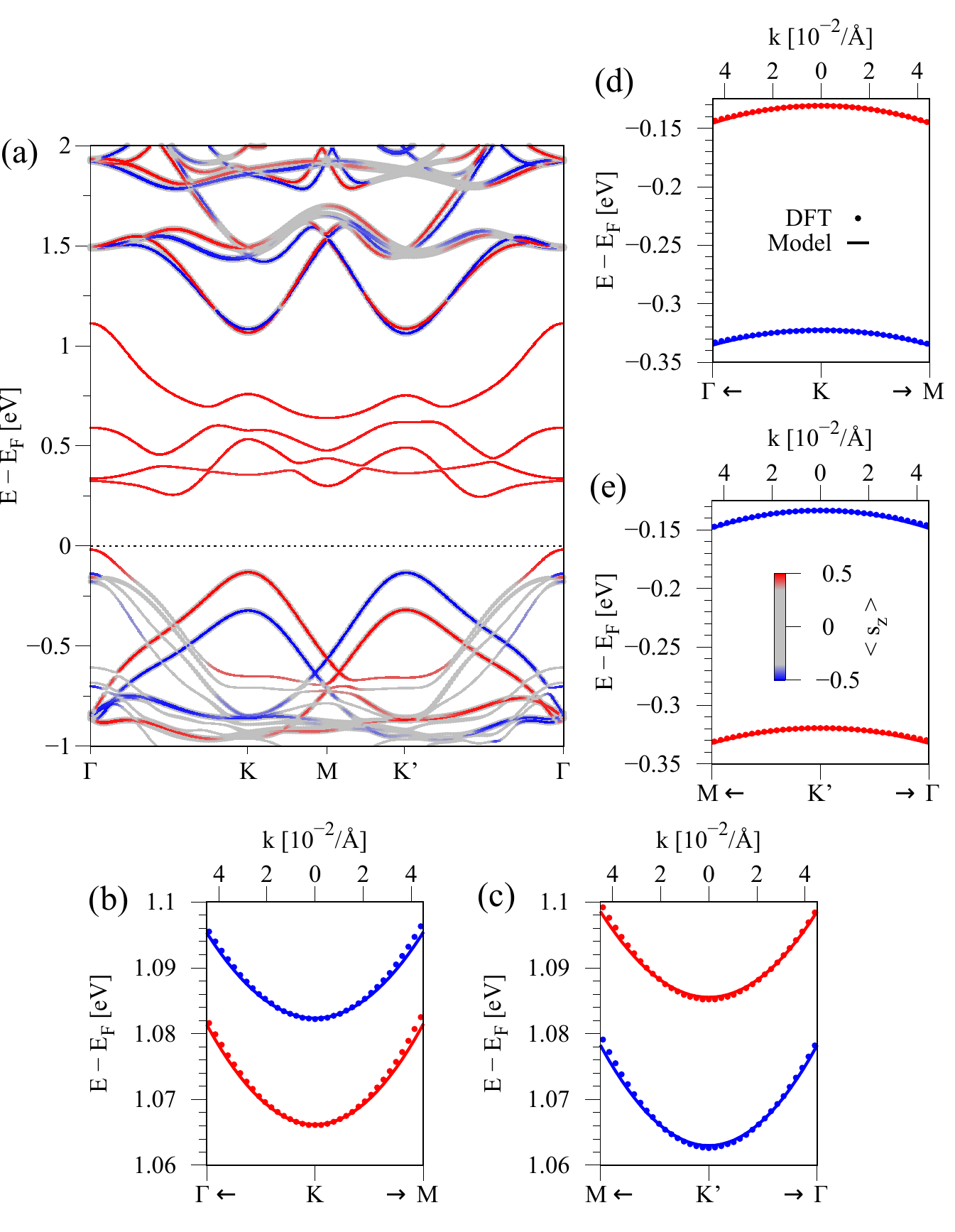}
 \caption{(Color online) Calculated band structure of MoSe$_2$/CrI$_3$ with SOC for the
 $0^{\circ}$ twist angle structure. 
 (a) Band structure along high symmetry lines. Color corresponds to the $s_z$-expectation value.
 (b,c) Zoom to the CB edge corresponding to MoSe$_2$ at K and K'. 
 Symbols are DFT data and solid lines are the fitted model Hamiltonian. 
 (d,e) Same as (b,c) but for VB edge.
 }\label{Fig:bands_TMDC_TMTH_SP_SOC}
\end{figure}

As a magnetic field or proximity exchange breaks time-reversal symmetry, we show the calculated
band structure for MoSe$_2$/CrI$_3$ with SOC in Fig. \ref{Fig:bands_TMDC_TMTH_SP_SOC}, for the $0^{\circ}$ twist angle supercell.
We find a very good agreement between the model Hamiltonian and the calculated bands around K and K' valley.
The valley degeneracy is now broken, especially when comparing the CB edges
at K and K' valley, see Figs. \ref{Fig:bands_TMDC_TMTH_SP_SOC}(b,c). 
Therefore, a TMDC/CrI$_3$ heterostructure shows valley polarization of the TMDC, 
in agreement with recent measurements \cite{Seyler2018:NL, Zhong2017:SA}, 
and other first-principles calculations \cite{Lin2019:ACS}. 
Inclusion of SOC effects for the large supercell structure is beyond our computational 
possibilites and therefore not included here.

\begin{table*}[!htb]
\begin{ruledtabular}
\begin{tabular}{l  c  c c  c  c   c  c  c}
 &\multicolumn{3}{c}{MoSe$_2$/CrI$_3$} &\multicolumn{3}{c}{WSe$_2$/CrI$_3$} 
 & MoSe$_2$/CrI$_3$ & WSe$_2$/CrI$_3$ \\
 \hline
   dipole [Debye] & \multicolumn{3}{c}{0.103}&\multicolumn{3}{c}{0.156} & 0.172 &  0.790 \\
 distance [\AA] & \multicolumn{3}{c}{3.506} &\multicolumn{3}{c}{3.497}  & 3.517 & 3.529 \\
 twist angle [$^{\circ}$] & \multicolumn{3}{c}{0} & \multicolumn{3}{c}{0} & 30 & 30 \\
 \hline
 calculation& B & noSOC & SOC& B & noSOC & SOC & noSOC & noSOC \\
 \hline
 $\Delta$ [eV] & 1.302 & 1.305 & 1.301 & 1.327 & 1.358 & 1.327 & 1.351 & 1.417\\
 $v_{\textrm{F}}$ [$10^{5} {\textrm{m}}/{\textrm{s}}$] &4.591& 4.579 & 4.591 & 5.863& 5.799 & 5.845 & 4.597 & 5.863\\
  $\lambda_{\textrm{c}}$ [meV]  &-9.647 & - & -9.678 & 13.90 & -& 13.81 & - & - \\
 $\lambda_{\textrm{v}}$ [meV]  & 94.56 & - & 94.43 & 241.79 & -&240.99 & - & - \\
 $\lambda_{\textrm{R}}$ [meV]  & - & - & - & - & -&  - & - & - \\
 $B_{\textrm{c}}$ [meV]  & - & -2.081 & -1.592 & - &-2.223 & -1.783 & -1.641 & -1.648\\
 $B_{\textrm{v}}$ [meV]  & - & -1.454 & -1.426 & - &-1.446 &-1.583 & 0.502 & 1.896 \\
\end{tabular}
\end{ruledtabular}
\caption{\label{Tab:fit_TMDC_CrI3} Summary of fit parameters, calculated dipoles 
and distances for TMDC/CrI$_3$ systems. 
The orbital gap $\Delta$ of the spectrum and the 
Fermi velocity $v_{\textrm{F}}$. The parameters $\lambda_{\textrm{c}}$ and $\lambda_{\textrm{v}}$ describe the SOC
splittings, and $B_{\textrm{c}}$ and $B_{\textrm{v}}$ 
are the proximity exchange parameters for CB and VB.
The dipole of the structures is given in debye and the distance d$_{\textrm{TMDC/CrI$_3$}}$ 
is defined in Fig. \ref{Fig:Figure1}(a). 
The twist angle $0^{\circ}$ ($30^{\circ}$) corresponds to the small (large) supercell in Fig. \ref{Fig:Figure1}. 
The individual columns are for calculations of the bare TMDC (B) with the modified lattice 
constant from the heterostructure, the heterostructure without SOC (noSOC) and with SOC. 
}
\end{table*}

In order to qualitatively extract the influence from the FM substrate, 
we calculate the band structure 
of the TMDC/CrI$_3$ heterostructures, for MoSe$_2$ and WSe$_2$ in the small supercell 
geometry, and fit it to our low energy Hamiltonian \footnotemark[1]. 
To obtain reasonable fit parameters we consider three situations. 
First, we calculate the band structure for the 
bare TMDC, removing the CrI$_3$, and check the SOC parameters for this situation, 
with the modified lattice constant used in the heterostructure geometry.
This is crucial, because an increase in the lattice constant diminishes for example 
the band gap of the TMDC \cite{Fang2018:PRB, Lloyd2016:NL, Frisenda2017:2DMA, Plechinger2015:2DM}. 
Second, we calculate the band structure for the TMDC/CrI$_3$ heterostructures without SOC in order
to obtain the proximity exchange splitting parameters. 
Finally, we calculate the band structure for the TMDC/CrI$_3$ heterostructures with SOC. 
In Tab. \ref{Tab:fit_TMDC_CrI3} we summarize the fit results for the three mentioned calculations, for $0^{\circ}$ twist angle. 
Additionally, we give the fit results for the $30^{\circ}$ twist angle case, corresponding to Figs. \ref{Fig:Figure1}(f-i) without SOC.

Especially interesting are the proximity exchange parameters, being roughly $2$~meV in magnitude, 
translating into about 10~T exchange field \cite{Srivastava2015:NP, Aivazian2015:NP, Li2014i:PRL, MacNeill2015:PRL}, 
in agreement with recent experiments \cite{Zhong2017:SA, Seyler2018:NL}.
The calculation of the atomic magnetic moments reveals, that the magnetization 
direction of the TMDC is 
the same as in the I atoms, opposite to the Cr atoms, therefore giving \textit{negative} proximity exchange parameters for
a net CrI$_3$ magnetization pointing along positive $z$-direction towards the TMDC. 
As there are many atoms in the unit cells, we calculate averaged magnetic moments for the different atomic species.
The averaged magnetic moments for the small non-twisted supercell are:
Cr ($+3.53~\mu_{B}$), I ($-0.19~\mu_{B}$), Mo ($-0.0039~\mu_{B}$), and Se ($-0.0046~\mu_{B}$). We averaged only over Se atoms closer to the CrI$_3$ substrate, as they mediate the proximity exchange to the Mo atoms.
The averaged magnetic moments for the large twisted supercell are: Cr ($+3.46~\mu_{B}$), I ($-0.18~\mu_{B}$), 
Mo ($-0.0029~\mu_{B}$), and Se ($-0.0031~\mu_{B}$).

Including SOC in the heterostructure calculations, we find the parameters to be barely different than those from the individual 
calculations for the bare TMDCs or for the heterostructures without SOC. 
Therefore, combining the model Hamiltonian, 
the SOC parameters from the bare TMDC monolayer, and the exchange parameters from 
the heterostructure calculation without SOC, already suffices to get a reasonable low energy band structure.
Our fit shows that the Rashba parameter is not necessary to 
capture the essentials of the band structure for the TMDC/CrI$_3$ stacks, because 
there is no in-plane component of the spin expectation value present around the band edges.

From the experimental point of view, when materials are mechanically exfoliated 
and stacked, one can expect various local interface (stacking) configurations between the TMDC and CrI$_3$, different to what is shown in Fig.~\ref{Fig:Figure1}(b). 
However, our calculations show, that proximity exchange splittings are marginally affected by different interface configurations, see Supplemental Material \footnotemark[1].

\section{Short-rangeness of proximity effects}

\begin{figure*}[!htb]
 \includegraphics[width=.99\textwidth]{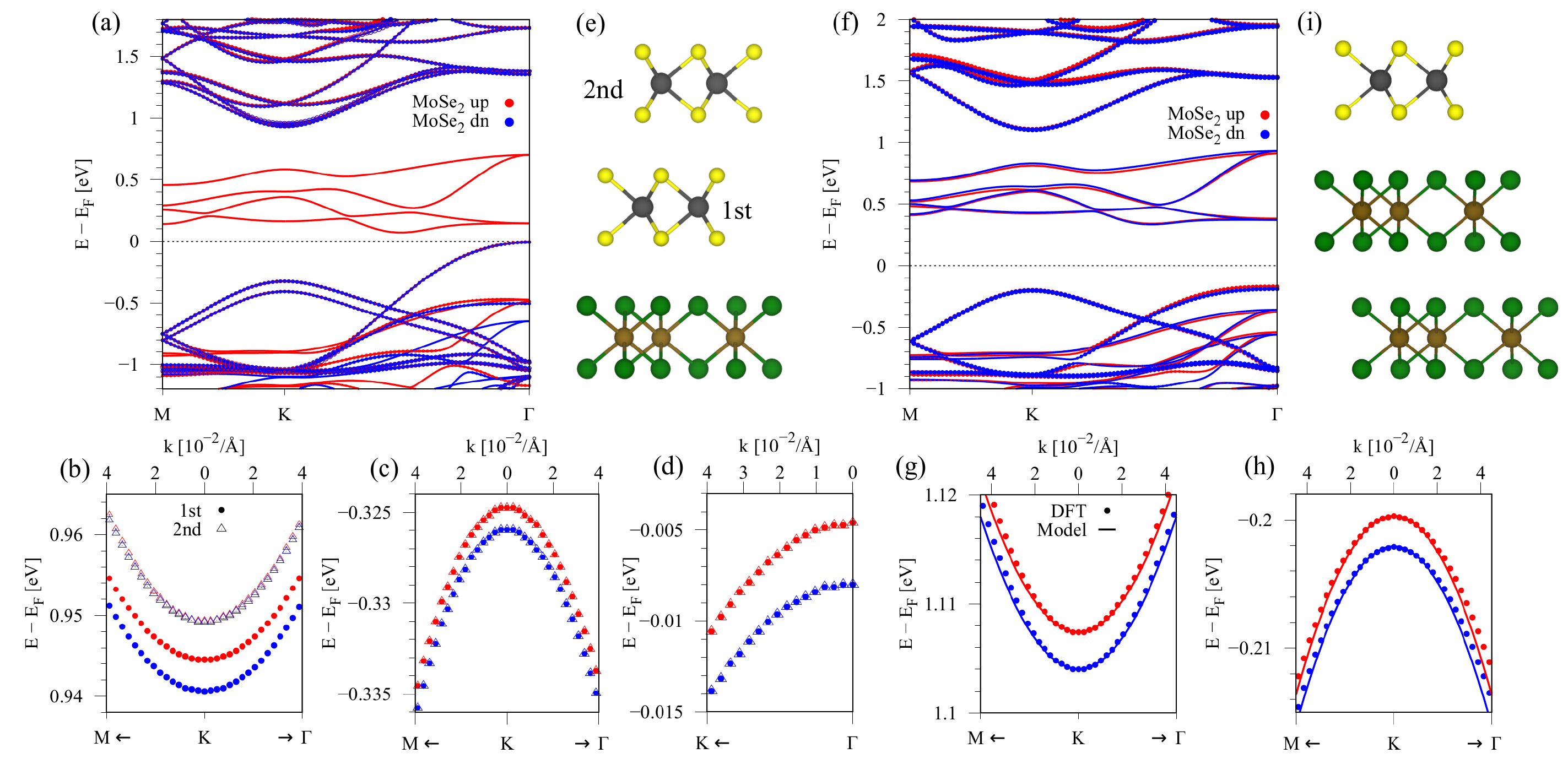}
 \caption{(Color online) 
 (a) Band structure along high symmetry lines for bilayer MoSe$_2$ on top of monolayer CrI$_3$. 
 In all subfigures (a-h), spin up (down) bands are plotted in red (blue), while symbols emphasize 
 the character of the bands. In (a) the bands corresponding to bilayer MoSe$_2$ are
 emphasized by thicker spheres. 
 (b) Zoom to the CB edge around the K point. (c,d) Same as (b), but for VB edge around K and $\Gamma$ point. 
 In (b-d) the bands corresponding to the first (second) layer of the bilayer MoSe$_2$ are
 emphasized by filled spheres (open triangles). 
 (e) Side view of the geometry with definition of first and second layer of 2H bilayer MoSe$_2$.
 (f) Band structure along high symmetry lines for monolayer MoSe$_2$ on top of bilayer CrI$_3$ in the AFM ($\uparrow\downarrow$) configuration.
 (g) Zoom to the CB edge around the K point (symbols), with a fit to the model Hamiltonian (solid line).
 (h) Same as (g), but for VB edge. 
 (i) Side view of the geometry of monolayer MoSe$_2$ on bilayer CrI$_3$.
 }\label{Fig:2xTMDC_CrI3_bands}
\end{figure*}

\subsection{\texorpdfstring{Bilayer TMDC on monolayer CrI$_3$}{}}
In experiment, when the TMDC is exfoliated from bulk crystals, 
it may also happen that a bilayer TMDC is transferred onto the magnetic CrI$_3$ substrate.
Does the second TMDC layer also experience proximity exchange, or only the closest layer?
To answer this question, we consider 2H bilayer MoSe$_2$ on top of monolayer CrI$_3$.
We first allow for relaxation of the whole geometry, similar to the monolayer MoSe$_2$ case, to get reasonable interlayer distances.

In Fig. \ref{Fig:2xTMDC_CrI3_bands}(a) we show the calculated band structure of 
2H bilayer MoSe$_2$ on top of monolayer CrI$_3$ without considering SOC effects. 
The bilayer TMDC has an indirect band gap, 
consistent with literature \cite{Kormanyos2018:PRB, Bussolotti2018:NF}.
For the CB edge, see Fig. \ref{Fig:2xTMDC_CrI3_bands}(b), we can identify the bands originating from the first and second layer of the bilayer MoSe$_2$. 
The bands from the first layer are spin-split, similar to
what we observe for the monolayer MoSe$_2$ case, while
the bands originating from the second layer experience negligible splitting.
The VB edge around the K point, see Fig. \ref{Fig:2xTMDC_CrI3_bands}(c), is formed by two non-degenerate bands, a spin up and a spin down one, which are split in energy. The magnitude of the splitting is again similar to the monolayer case and a result of proximity exchange. However, we find that each band is formed by orbitals from both TMDC layers, which does not allow us to trace back the splitting to proximity exchange in a specific layer. 
For further insights, we show the calculated layer- and spin-resolved density of states, see Supplemental Material \footnotemark[1]. In additon, the calculated magnetic moments for the second 
layer ($-0.0001~\mu_B$) are much 
smaller than for the first layer ($-0.004~\mu_B$). 
Thus we conclude that the proximity exchange is mainly 
induced in the TMDC layer closest to the CrI$_3$ substrate. 

\subsection{\texorpdfstring{TMDC on antiferromagnetic bilayer CrI$_3$}{}}
Another feature which was especially observed in bilayer CrI$_3$ is the switching from FM
to AFM coupled layers by gating \cite{Huang2017:Nat, Huang2018:NN}. 
For the AFM state, there are two energetically degenerate states of the bilayer
(first layer $\uparrow$, second layer $\downarrow$, or vice versa). 
Therefore, we calculate the band structure, with and without SOC, for MoSe$_2$ on top of 
bilayer CrI$_3$, which is stacked in the low temperature 
rhombohedral phase \cite{Jiang2018:arxiv, Soriano2018:arxiv, Sivadas2018:NL}. 
We find a total energy difference of $E_{\textrm{AFM}}-E_{\textrm{FM}} \approx 24$~meV 
between the FM and the AFM state calculated without SOC 
of the whole TMDC/bilayer-CrI$_3$ stack, which contains four Cr atoms in our supercell.  
In agreement with previous DFT calculations \cite{Soriano2018:arxiv,Jiang2018:arxiv, Sivadas2018:NL}, 
we find that
the FM state of the bilayer CrI$_3$ is energetically 
favorable compared to the AFM state 
in contrast to experiments \cite{Huang2017:Nat, Jiang2018:NM, Song2018:SC}.

In Fig. \ref{Fig:2xTMDC_CrI3_bands}(f) we show the calculated band structure of 
MoSe$_2$ on top of bilayer CrI$_3$ without SOC, 
when the bilayer CrI$_3$ is in the AFM ($\uparrow\downarrow$) configuration (the magnetization of the CrI$_3$ layer directly below the TMDC is pointing $\uparrow$).
In the Supplemental Material \footnotemark[1], we show the band structures including SOC for both cases, the FM and AFM configuration. 
The fit to the low energy bands, see Figs. \ref{Fig:2xTMDC_CrI3_bands}(g,h) are similar 
to what is shown in Fig. \ref{Fig:Figure1} 
for MoSe$_2$ on top of monolayer CrI$_3$.
The naive expectation is that, depending on the total magnetization 
of the bilayer CrI$_3$, we can enhance or reduce proximity exchange in the 
TMDC, compared to the monolayer CrI$_3$ case. However, we find that the 
FM~($\uparrow\uparrow$) or AFM ($\uparrow\downarrow$) coupled bilayer give almost no difference 
in the fit parameters compared to the monolayer case \footnotemark[1].

\subsection{Effects of hBN barrier}
Finally, when introducing a hBN buffer layer between the MoSe$_2$ and CrI$_3$, proximity 
exchange splittings of the MoSe$_2$ bands are drastically reduced to about $100~\mu$eV \footnotemark[1].
The calculated averaged magnetic moments for the case of MoSe$_2$/hBN/CrI$_3$ are: Cr ($+3.50~\mu_{B}$), I ($-0.19~\mu_{B}$), N ($-0.0016~\mu_{B}$), B ($0.0~\mu_{B}$), Mo ($-0.0001~\mu_{B}$), and Se ($-0.0001~\mu_{B}$). The magnetic moments in the TMDC are reduced by one order of magnitude, 
compared to the case without the hBN layer.
By looking at the proximity exchange in the hBN layer, we find that the bands originating from hBN are strongly hybridized with the CrI$_3$ bands, see Supplemental Material \footnotemark[1]. 
This can be helpful for interpreting tunneling experiments of such heterostructures.
All these results indicate that proximity exchange is truly a short range effect, and 
can be used to create and detect the magnetic order in the layered AFM, bilayer CrI$_3$ \footnotemark[1].

\section{Gate tunable proximity exchange and exciton splitting}
Motivated by recent experiments \cite{Jiang2018:NN, Huang2018:NN, Huang2017:Nat}, 
showing the electric field control of magnetism in few layer CrI$_3$,
and the optical tuning of proximity exchange in TMDC/CrI$_3$ heterostructures \cite{Seyler2018:NL}, we perform
additional calculations for our heterostructures, where we apply a transverse electric field across the geometry consisting of one monolayer of TMDC and one of CrI$_3$.

\begin{figure}[!htb]
 \includegraphics[width=.99\columnwidth]{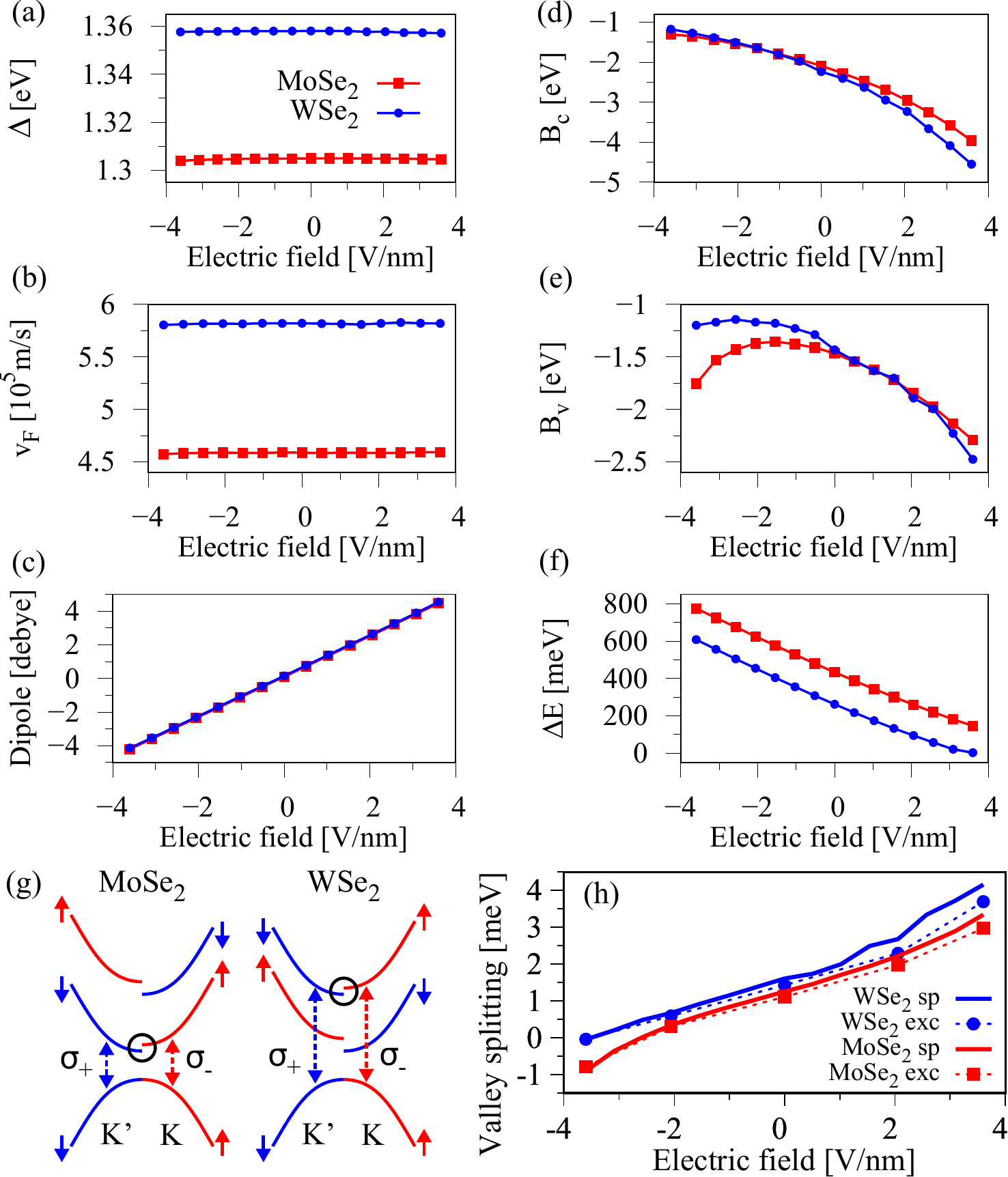}
 \caption{(Color online) Fit parameters as a function of transverse electric field
 for TMDC/CrI$_3$ heterostructures for calculations without SOC. (a) The orbital gap $\Delta$ of the spectrum, (b) the 
Fermi velocity $v_{\textrm{F}}$, (c) the dipole of the heterostructure, 
(d,e) the proximity exchange parameters $B_{\textrm{c}}$ and $B_{\textrm{v}}$ 
and (f) the band gap $\Delta$E, as defined in Fig. \ref{Fig:Figure1}(c). 
(g) Schematic representation of the exchange interaction 
 in the band structure of MoSe$_2$ and WSe$_2$ at K and K' points depicting the lowest energy 
 optically allowed transitions (vertical dashed arrows). (h) Single-particle and intralayer exciton valley splitting for the first active optical transition between 
 the top VB and the first (second) CB in MoSe$_2$ (WSe$_2$) 
 [indicated by the circle in (g)] as function of the external electric field. For the excitons, the valley splitting is taken as the energy difference between the first absorption peaks (see Fig.~\ref{Fig:Figure1} and Supplemental Material).
 }\label{Fig:Efield_TMDC_TMTH}
\end{figure}
As calculations without SOC already give reasonable proximity exchange parameters, we 
neglect SOC for the electric field study.
In Fig. \ref{Fig:Efield_TMDC_TMTH} we show the fit parameters as a function of 
a transverse electric field for TMDC/CrI$_3$ heterostructures calculated without SOC, for the $0^{\circ}$ twist angle structures. 
We find that the gap parameter $\Delta$, as well as the Fermi velocity 
$v_{\textrm{F}}$ are barely affected by external electric
fields. The dipole of the heterostructure depends linearly on the electric field.  
By applying an electric field, the band offsets can be changed. 
The band gap $\Delta$E of the heterostructure, defined in Fig. \ref{Fig:Figure1}(c), 
shrinks linearly with applied electric field. 
This tunability of the band offsets could be very important for other effects. 
Imagine electrons located in the CrI$_3$ layer coupled to holes in the TMDC layer. 
As we apply an electric field, we tune the band gap $\Delta$E, 
possibly affecting the lifetime of interlayer excitons.
Especially interesting is the fact that the CB states, originating from the CrI$_3$, 
are spin polarized, see Fig. \ref{Fig:Figure1}(c), which then gives
additional valley control, depending on the magnetization of the CrI$_3$, due to spin-valley coupling in the TMDC.
Most important, the two exchange parameters $B_{\textrm{c}}$ and $B_{\textrm{v}}$ can be tuned by the external electric field. 
In general, the proximity exchange increases, when the electric field is tuned from negative to positive values,
which enables the gate control of proximity exchange. 

Let us now look at the tunability of proximity exchange 
reflected in the valley splitting of the TMDC intralayer excitons. 
In Fig.~\ref{Fig:Efield_TMDC_TMTH}(g) we sketch the energy levels for the top VB 
and first two CBs at K and K' valleys for MoSe$_2$ and WSe$_2$. We set the 
top VB to zero, which simplifies the analysis by just looking at the allowed 
optical transitions that satisfy the spin-valley locking. 
The effective signature of proximity is the interplay of SOC 
and exchange parameters in the optically allowed transitions from the VB to 
the first (second) CB in MoSe$_2$ (WSe$_2$). The resulting valley splittings  
are shown in Fig.~\ref{Fig:Efield_TMDC_TMTH}(h) for the first exciton absorption peak (dashed lines 
with points) and the single-particle of the optically active bands (solid lines). Unlike the 
strong nonlinear behavior observed in the exchange parameters $B_\text{c}$ and $B_\text{v}$ under 
applied electric field, see Figs.~\ref{Fig:Efield_TMDC_TMTH}(d-e), the optical valley splitting due the 
proximity exchange shows a rather linear behavior, with the single-particle results following closely 
the excitonic calculations. Specifically for MoSe$_2$, the valley splitting changes sign at an electric field of about -2.5 V/nm. This trend might also happen for WSe$_2$ for further negative values 
of electric field. Therefore, in addition to the control of the twist angle, 
see Figs.~\ref{Fig:Figure1}(j,k), the application of external electric fields can modify the 
value of the valley splitting. In real samples it is reasonable to expect an interplay of 
both effects, the twist angle and the electric field.  Regarding the parameters used in the calculations shown 
in Fig.~\ref{Fig:Efield_TMDC_TMTH}(h), we used the model Hamiltonian with the values of 
$\Delta$, $v_{\text{F}}$, $\lambda_\text{c}$, $\lambda_\text{v}$ given in 
Tab.~\ref{Tab:fit_TMDC_CrI3} with SOC and $B_\text{c}$, $B_\text{v}$ extracted from 
the data presented in Figs.~\ref{Fig:Efield_TMDC_TMTH}(d,e). See Supplemental Material \footnotemark[1] 
for details and the calculated absorption spectra used to extract the exciton valley splittings.

Finally, we want to discuss several experimental uncertainties, one has to consider, 
before directly comparing them with our results. For example, the twist angle and a resulting moir\`{e} pattern between the TMDC and CrI$_3$ in micrometer size samples, can influence proximity exchange. 
We have seen, that the twist angle can lead to a giant enhancement of the valley splitting. 
However, when twist angle is not an issue and can be precisely controlled in experiment, 
there are still several different local interface configurations. We have studied this as 
different interface geometries of the small supercell in the Supplemental Material \footnotemark[1]. 
Local variations in the magnitude of the proximity exchange and valley splitting can occur. 
One can even speculate about a vanishing global proximity exchange in the TMDC, when thinking of magnetic domains in the CrI$_3$ substrate.
We have also seen, that proximiy exchange depends on the actual electric dipole field across the sample.
In contrast to our approach of mono- and bilayer CrI$_3$ as substrate, experiments may utilize thicker CrI$_3$ samples (few layers) affecting the overall electrostatics and the band alignment in the heterostructure. In this context, one must also consider the effect of an additional SiO$_2$ or hBN substrate/capping layer to protect the system from the environment.  
In additon, recent first-principles calculations have shown that the optical and magneto-optical properties of CrI$_3$ are also dominated by strongly bound excitons \cite{Wu2019:NC}.
When studying the absorption spectra of TMDC/CrI$_3$ heterostructures, one has to be aware that quasiparticles can in principle be created in both layers simultaneously.
We conclude that experimentally (and also theoretically) one has to be very careful in preparing the heterostructures and analyzing the data, in order to make qualitative statements 
about the proximity exchange effects. 
Even though our presented analysis is very systematic, we can at most give predictions for idealized structures.

\section{Summary}
By combining DFT calculations with a low energy model Hamiltonian of exchange proximitized TMDCs,
we have shown that a CrI$_3$ substrate causes sizable proximity exchange in the TMDCs
MoSe$_2$ and WSe$_2$. 
Crucial for the magnitude of the induced valley splitting, is the twist angle between the TMDC and CrI$_3$, as we find from the $0^{\circ}$ and $30^{\circ}$ twist angle cases, by calculating optical absorption spectra.
By applying experimentally accessible electric fields transverse to the heterostructure, 
we can tune band offsets, proximity exchange, and consequently the valley splitting in the TMDCs. 
Finally, we have seen that proximity exchange originates only from the FM 
CrI$_3$-layer closest to the TMDC by investigating TMDC/bilayer-CrI$_3$ heterostructures. 
The observed twist angle dependence, electric field tunability, and short-rangeness of proximity exchange
are experimentally testable fingerprints of our results, and should be generally valid for other
2D van der Waals heterostructures.

\acknowledgments
This work was supported by DFG SPP 1666, SFB 1277 (B05, B07), the European Unions Horizon 2020 research and innovation program under Grant No. 785219, the Alexander von Humboldt Foundation and Capes (grant No. 99999.000420/2016-06).

\footnotetext[1]{See Supplemental Material at [URL will be inserted by publisher], including Refs. \cite{ASE, Jiang2018:NL, McGuire2015:CoM, Schutte1987:JSSC, James1963:AC, Song2018:NL, Hohenberg1964:PRB, Giannozzi2009:JPCM, Kresse1999:PRB, Perdew1996:PRL, Liechtenstein1995:PRB, Grimme2006:JCC,Barone2009:JCC, Bengtsson1999:PRB, Qi2015:PRB, Kormanyos2014:2DM, RohlfingPRB:2000,Scharf2017:PRL,Tedeschi2019:PRB,FariaJunior2019phosphorene, Rytova1967coulomb,Keldysh1979coulomb,Cudazzo2011:PRB,Berkelbach2013:PRB, Huang2017:Nat, Berkelbach2013:PRB, Scharf2017:PRL, Haug2009book, Scharf2019:JPCM, Zhang2019:PRB, Seyler2018:NL, Catellani1987:PRB, Huang2018:NN, Jiang2018:NM,Jiang2018:NN, Dirnberger2018:arxiv, Zhong2017:SA, McGuire2017:C, Webster2018:PRB, Zhang2015:JMCC, Handy1952:JACS, Wakabayashi1975:PRB, Webster2018:PCCP,Liu2016:PCCP,Lado2017:2DM, Jiang2019:PLA, KumarGudelli2019:NJP, Behera2019:APL}, for details about the used supercell geometries, the computational methods, model Hamiltonian, and futher calculation results.}

\bibliography{paper}

\cleardoublepage


\section*{Supplementary material}

# Supplemental Material:
# Proximity exchange effects in MoSe$_2$ and WSe$_2$ heterostructures with CrI$_3$: twist angle, layer, and gate dependence

Klaus Zollner,[1, *] Paulo E. Faria Junior,[1] and Jaroslav Fabian[1]

[1]*Institute for Theoretical Physics, University of Regensburg, 93040 Regensburg, Germany*

The Supplemental Material contains details about the used supercell geometries, the computational methods, as well as the model Hamiltonian used to fit the low energy bands of the TMDCs in the presence of proximity exchange. Further, we describe the computational details for the intralayer exciton calculations and show absorption spectra for MoSe$_2$ and WSe$_2$ for a series of transverse electric fields. We also include a brief discussion about different interface configurations of MoSe$_2$/CrI$_3$, the effect of the Hubbard $U$ parameter, summarize fit parameters for the bilayer CrI$_3$ cases, show the spin-resolved density of states for bilayer-TMDC on CrI$_3$, and show results for a MoSe$_2$/hBN/CrI$_3$ trilayer structure. Finally, we propose device geometries for the creation and detection of magnetic order in bilayer CrI$_3$.

## S1. GEOMETRY AND COMPUTATIONAL DETAILS

To study proximity exchange effects, we set up a common unit cell for the TMDC/CrI$_3$ heterostructure with the atomic simulation environment (ASE)[1]. We choose a 2×2 supercell for the TMDCs (MoSe$_2$ and WSe$_2$) and a 1×1 cell of CrI$_3$. The lattice constants and strains are summarized in Tab. S1, for the individual monolayers. We can see that a maximum strain of roughly 3% is present for CrI$_3$, which is still an acceptable value for studying heterostructures, without altering the individual monolayer properties too much. Taking the geometry as explained, we end up with 20 atoms, which is our small supercell geometry with 0° twist angle. Our heterostructure is built such that a chalcogen atom of the TMDC is directly above a Cr atom of the CrI$_3$ layer (see Fig. 1 in the main manuscript). Other stacking configurations, can give slight variations in the band alignment and spin splittings. For proper interlayer distances, we allow the atoms of the TMDC, as well as the Cr atoms in the CrI$_3$, to relax their z coordinates, while the I atoms are allowed to fully change their position in x, y, and z, because they form a distorted octahedral surrounding around the Cr-atoms[2]. The distances $d_{\text{TMDC/CrI}_3}$ are about 3.5 Å, and of typical size for van der Waals systems.

|  | CrI$_3$ | MoSe$_2$ | WSe$_2$ |
| --- | --- | --- | --- |
| $a$ (exp.) [Å] | 6.867 | 3.288 | 3.282 |
| $a$ (het.) [Å] | 6.748 | 3.374 | 3.374 |
| strain [%] | -1.73 | 2.62 | 2.80 |

TABLE S1. Lattice constants and strains for the subsystems used in the TMDC/CrI$_3$ heterostructures. The experimental $a$ (exp.) lattice constants (Refs.[3–5]) and lattice constants used for the heterostructures $a$ (het.) are given. The strain for each subsystem, is calculated as $(a_{\text{het}} - a_{\text{exp}})/a_{\text{exp}}$.

In addition, we consider a larger 7 × 7 supercell of MoSe$_2$ on top of a $(2\sqrt{3} \times 2\sqrt{3})$R30° supercell of CrI$_3$. We change the lattice constant of MoSe$_2$[5] to $a = 3.358$ Å and the one of CrI$_3$[3] to $a = 6.785$ Å, resulting in strains of 2.13% and -1.19%. Similarly, we also consider a larger supercell structure for WSe$_2$ on CrI$_3$. These geometries then contain 243 atoms in the supercell. The relative twist of 30° between the two layers should give further insight in the proximity effect, as layer alignment might be hard to control in experiment. Recently, it has been shown, that the alignment can be crucial to explain realistic proximity effects in graphene/topological insulator heterostructures[6].

Due to experimental works of WSe$_2$/CrI$_3$ heterostructures[7,8], we first focused on this kind of structure, additonally considering MoSe$_2$. Fortunately, the lattice constants of Se-based TMDCs and CrI$_3$ can be easily matched into a small supercell, as discussed above. Then, also the introduced strains are within an acceptable limit. There is actually another transition-metal trihalide, CrBr$_3$[9–11], showing similar magnetic ordering as CrI$_3$ and whose lattice constant $a = 6.26$ Å[12] would be more appropriate to interface it with S-based TMDCs, at least from the theoretical perspective. The lattice constants of MoS$_2$ and WS$_2$ are about $a = 3.15$ Å[4,13]. However, we do not expect any major differences for S-based TMDCs. Certainly, there can be minor differences in band alignments or the magnitude of proximity exchange, but we expect the overall physics to be the same. Therefore, we focus only on Se-based TMDCs.

The electronic structure calculations and structural relaxation of our heterostructures are performed by means of density functional theory[14] within Quantum ESPRESSO[15]. Self-consistent calculations are done with the $k$-point

sampling of $36 \times 36 \times 1$ for the small non-twised TMDC/CrI$_3$ heterostrucures ($6 \times 6 \times 1$ for the large supercell with 30° twist angle). We perform open shell calculations that provide the spin polarized ground state with magnetization pointing in z-direction. We use an energy cutoff for charge density of 500 Ry, and the kinetic energy cutoff for wavefunctions is 60 Ry for the scalar relativistic pseudopotential with the projector augmented wave method[16] with the Perdew-Burke-Ernzerhof exchange correlation functional[17]. In the cases where SOC is included, the fully relativistic version of the pseudopotentials are used. In addition we include the Hubbard correction for Cr atoms with $U = 4$ eV[18]. For the relaxation of the heterostructures, we add van der Waals corrections[19,20] and use quasi-newton algorithm based on trust radius procedure. In order to simulate quasi-2D systems a vacuum of at least 20 Å is used to avoid interactions between periodic images in our slab geometry. Dipole corrections[21] are also included to get correct band offsets and internal electric fields. Structural relaxations are performed until all components of all forces were reduced below $10^{-3}$ [Ry/$a_0$], where $a_0$ is the Bohr radius.

## S2. MODEL HAMILTONIAN

We want to describe proximity exchange effects that are due to the magnetic insulator substrate CrI$_3$. Following an earlier work in this field[22], we introduce a minimal model Hamiltonian to describe the band structure of the TMDC close to K and K', in the presence of proximity exchange

$$\mathcal{H} = \mathcal{H}_0 + \mathcal{H}_\Delta + \mathcal{H}_{\text{soc}} + \mathcal{H}_{\text{ex}} + \mathcal{H}_{\text{R}}, \tag{S1}$$

$$\mathcal{H}_0 = \hbar v_{\text{F}} s_0 \otimes (\tau \sigma_x k_x + \sigma_y k_y), \tag{S2}$$

$$\mathcal{H}_\Delta = \frac{\Delta}{2} s_0 \otimes \sigma_z, \tag{S3}$$

$$\mathcal{H}_{\text{soc}} = \tau s_z \otimes (\lambda_{\text{c}} \sigma_+ + \lambda_{\text{v}} \sigma_-), \tag{S4}$$

$$\mathcal{H}_{\text{ex}} = -s_z \otimes (B_{\text{c}} \sigma_+ + B_{\text{v}} \sigma_-), \tag{S5}$$

$$\mathcal{H}_{\text{R}} = \lambda_{\text{R}} (\tau s_y \otimes \sigma_x - s_x \otimes \sigma_y). \tag{S6}$$

The valley index is $\tau = \pm 1$ for K (K') point and $v_{\text{F}}$ is the Fermi velocity. The Cartesian components $k_x$ and $k_y$ of the electron wave vector are measured from K (K'). The pseudospin Pauli matrices are $\sigma_i$ acting on the (CB, VB) subspace and spin Pauli matrices are $s_i$ acting on the ($\uparrow,\downarrow$) subspace, with $i = \{0, x, y, z\}$. The parameter $\Delta$ denotes the orbital gap of the spectrum. For short notation we introduce $\sigma_\pm = \frac{1}{2}(\sigma_0 \pm \sigma_z)$. The parameters $\lambda_{\text{c}}$ and $\lambda_{\text{v}}$ describe the spin splitting for CB and VB, due to SOC, respectively. In the case when we have a ferromagnetic substrate, proximity exchange effects will be present with $B_{\text{c}}$ and $B_{\text{v}}$ describing the proximity induced exchange splittings of CB and VB. The Rashba SOC parameter $\lambda_{\text{R}}$ is due to the presence of inversion asymmetry in the heterostructure. The four basis states we use are $|\Psi_{\text{CB}}, \uparrow\rangle$, $|\Psi_{\text{VB}}^\tau, \uparrow\rangle$, $|\Psi_{\text{CB}}, \downarrow\rangle$, and $|\Psi_{\text{VB}}^\tau, \downarrow\rangle$. The wave functions are $|\Psi_{\text{CB}}\rangle = |d_{z^2}\rangle$ and $|\Psi_{\text{VB}}^\tau\rangle = \frac{1}{\sqrt{2}}(|d_{x^2-y^2}\rangle + \text{i}\tau|d_{xy}\rangle)$, corresponding to CB and VB at K and K', since the band edges are formed by different $d$-orbitals from the transiton metal[23] of the TMDC.

## S3. INTRALAYER EXCITONS

To investigate the excitonic effects within the TMDC layer, the so-called intralayer excitons, we employ the robust formalism of the effective Bethe-Salpeter equation (BSE)[24–27]. Focusing on direct intralayer excitons at zero temperature and without doping effects, the BSE is essentially a typical eigenvalue problem

$$\left[E_c(\vec{k}) - E_v(\vec{k}) - \Omega_N\right] A_{c,v,\vec{k}}(N) + \sum_{c'v'\vec{k}'} \mathbb{D}_{c'v'\vec{k}'}^{cv\vec{k}} A_{c',v',\vec{k}'}(N) = 0, \tag{S7}$$

with eigenvalues being the exciton energies, $\Omega_N$ (with $N$ labeling the excitonic states), and eigenvectors being the exciton envelope functions, $A_{c,v,\vec{k}}(N)$, defined in terms of the total many-body wavefunction

$$\Psi_N(\vec{x}, \vec{x}') = \sum_{c,v,\vec{k}} A_{c,v,\vec{k}}(N) \psi_{c,\vec{k}}(\vec{x}) \psi_{v,\vec{k}}^*(\vec{x}'), \tag{S8}$$

in which the $\vec{x}$ includes spatial and spin degrees of freedom. The single-particle energies and wavefunctions for conduction and valence bands are the solutions obtained using the model Hamiltonian discussed in the previous section.

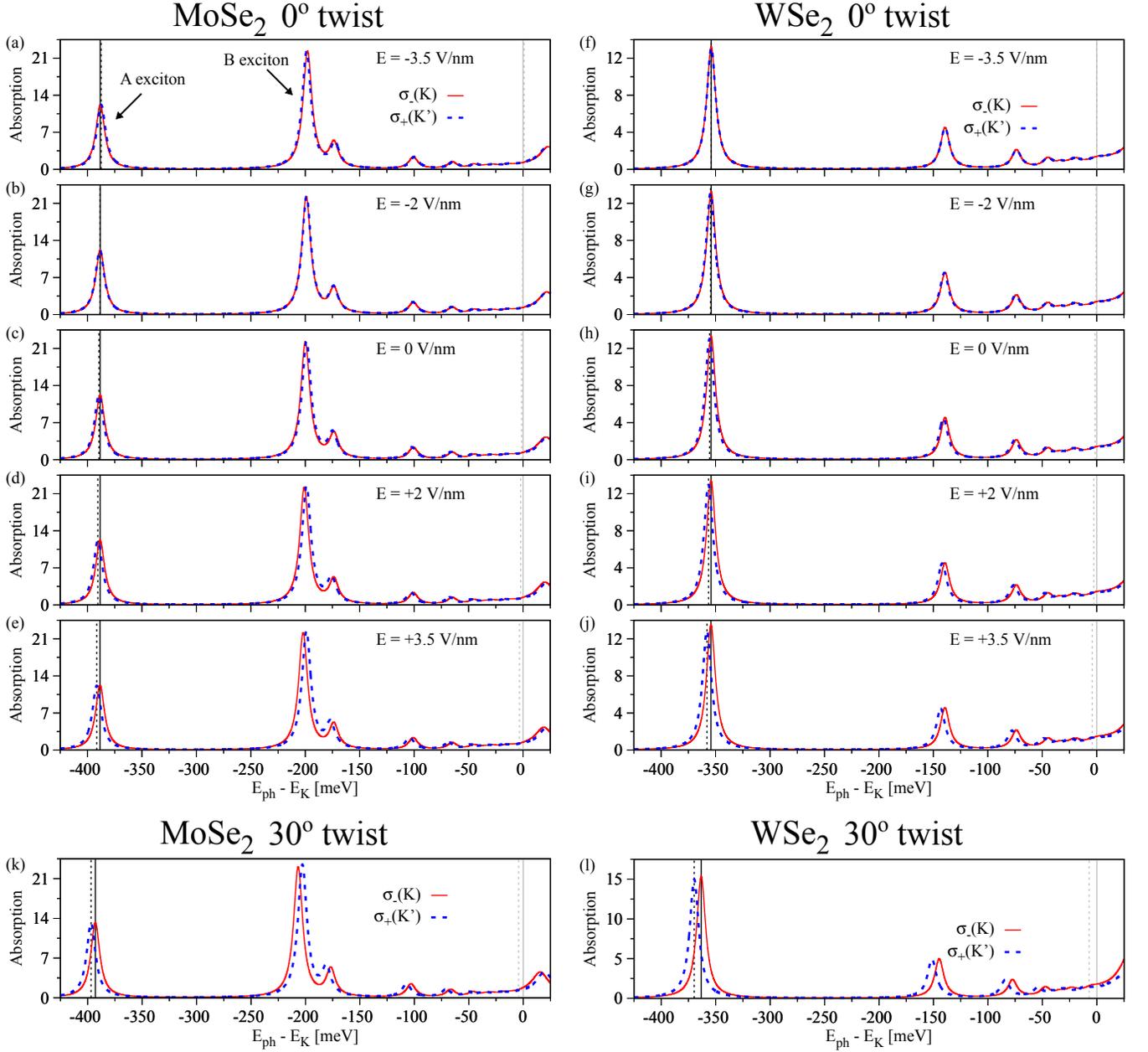

FIG. S1. (Color online) Calculated absorption spectra of intralayer excitons for (a-e) MoSe$_2$/CrI$_3$ and (f-j) WSe$_2$/CrI$_3$ with 0° twist angle for different values of external electric field. (k) MoSe$_2$/CrI$_3$ and (l) WSe$_2$/CrI$_3$ with 30° twist angle. The vertical solid (dashed) arrows indicate the energy contribution at K (K') point.

The direct electron-hole Coulomb term in the BSE is written as

$$\mathbb{D}^{cv\vec{k}}_{c'v'\vec{k}'} = -\int d\vec{x} \int d\vec{x}'\, \psi^*_{c,\vec{k}}(\vec{x})\psi_{v,\vec{k}}(\vec{x}')\, \mathrm{v}(\vec{r},\vec{r}')\, \psi_{c',\vec{k}'}(\vec{x})\psi^*_{v',\vec{k}'}(\vec{x}')\,, \tag{S9}$$

with the electron-hole interaction mediated by the Rytova-Keldysh potential[28–31], given by

$$\mathrm{v}(\rho) = \frac{e^2}{8\varepsilon_0 r_0}\left[ H_0\!\left(\frac{\varepsilon}{r_0}\rho\right) - Y_0\!\left(\frac{\varepsilon}{r_0}\rho\right) \right], \tag{S10}$$

in which $H_0$ is the zeroth-order Struve function, $Y_0$ is the zeroth-order Bessel function of the second kind, $\rho = |\vec{r}-\vec{r}'| = \sqrt{(x-x')^2 + (y-y')^2}$, $e$ is the electron charge, $\varepsilon_0$ is the vacuum permittivity, $\varepsilon = (\varepsilon_t + \varepsilon_b)/2$ is the effective dielectric
3

constant given by the average of top, $\varepsilon_t$, and bottom, $\varepsilon_b$, dielectric constants, and $r_0$ is the screening length. We consider TMDC layers with vacuum on top ($\varepsilon_t = 1$) and with CrI$_3$ below (with dielectric constant $\varepsilon_b = 1.8$, estimated from the experimental study of Huang et al.[32]). For the screening lengths of MoSe$_2$ and WSe$_2$ we take $r_0 = 5.1$ nm and $r_0 = 4.5$ nm, respectively, based on the values provided in the study of Berkelbach, Hybertsen, and Reichman[31]. We solve the BSE numerically using a 2D $k$-grid with size -0.5 to 0.5 Å$^{-1}$ in both $k_x$ and $k_y$ directions with total discretization of $101 \times 101$ points with spacing of $\Delta k = 10^{-2}$ Å$^{-1}$. Furthermore, the Coulomb potential is averaged in the vicinity of each $k$-point, in a square region of $-\Delta k/2$ to $\Delta k/2$ discretized with an internal mesh of $101 \times 101$ points. These number of points are more than enough to guarantee the convergence of the exciton energies as suggested by Scharf et al.[25] that used $60 \times 60$ points for the $k$-grid with $100 \times 100$ internal points.

The absorption spectra taking into account the excitonic effects is written as

$$\alpha^a(\hbar\omega) = C_0 \sum_N \left| \sum_{cv\vec{k}} A_{cv\vec{k}}(N) p^a_{vc}(\vec{k}) \right|^2 \delta(\Omega_N - \hbar\omega) \tag{S11}$$

with the superindex $a$ indicating the polarization of the light, $C_0 = \left(4\pi^2 e^2\right)/\left(\varepsilon_0 c_l \omega \mathcal{A} \hbar^2\right)$, $c_l$ is the speed of light (the subindex $l$ was added to not be confused with the conduction band index $c$), $\mathcal{A}$ is the 2D unit area and the dipole matrix element written as $p^a_{nm}(\vec{k}) = \frac{\hbar}{m_0} \left\langle n, \vec{k} \left| \hat{e}_a \cdot \vec{p} \right| m, \vec{k} \right\rangle$. To the final absorption spectra we apply a lorentzian broadening with energy dependent full width at half-maximum[33,34]

$$\Gamma(\hbar\omega) = \Gamma_1 + \frac{\Gamma_2}{1 + e^{[(E_0 - \hbar\omega)/\Gamma_3]}} \tag{S12}$$

using $\Gamma_1 = \Gamma_2 = \Gamma_3 = 10$ meV and $E_0$ the single-particle energy at $K$-point for the first allowed optical transition.

In the main text we have shown in Fig. 4(h) the proximity induced valley splitting between the first absorption peaks that take place at K and K' points for different electric field values. For completeness, we show here in Figs. S1(a-j) the full absorption spectra for intralayer excitons in MoSe$_2$/CrI$_3$ and WSe$_2$/CrI$_3$ for the different values of electric field. Furthermore, we present in Figs. S1(k-l) the full absorption spectra for the MoSe$_2$/CrI$_3$ and WSe$_2$/CrI$_3$ heterostructures with 30° twist angle. For MoSe$_2$, the A and B exciton peaks are visible in the energy range we present in the figures and regarding the valley splitting of the B exciton, it has the same value as the A exciton but with opposite sign.

## S4. EXPERIMENTALLY RELEVANT CONSIDERATIONS

### A. Different Interface Configurations

In experiment, the interface of the TMDC and the CrI$_3$ is, in general, not perfectly aligned as in our theoretical approach. Therefore, several different interfacial stacking configurations are likely to occur, possibly influencing the magnitude of proximity exchange in the TMDC. To get a more qualitative picture of the proximity exchange splitting, we consider different stacking configurations that are useful to explain experimental results. We consider the $2 \times 2$ supercell for the MoSe$_2$ and a $1 \times 1$ cell of CrI$_3$, where we change the relative alignment between the two, resulting in various interface geometries.

The calculated band structures for the different alignments of the MoSe$_2$ above the CrI$_3$ barely differ from each other, see Fig. S2. The main difference is in the spin splittings and the corresponding exchange parameters $B_c$ and $B_v$, which are diminished by about 0.6 meV, compared to what is found for the alignment from the main text. The results are in agreement with a recent publication[35], showing stacking dependent proximity exchange in WSe$_2$/CrI$_3$ heterostructures.

### B. Effect of Hubbard $U$ parameter

We want to analyze the effect of the Hubbard $U$ parameter on the proximity exchange. According to other first-principles studies of CrI$_3$[36–41], $U$ values within a range of 1–4 eV are a reasonable choice. In Fig. S3 we show the fit parameters, as well as the average magnetic moments on Cr, I, Mo, Se atoms, as function of the Hubbard $U$ parameter. We find that the orbital fit parameters, $\Delta$ and $v_F$ barely depend on $U$. In contrast, the proximity exchange paramters, $B_c$ and $B_v$, depend linearly on $U$. We also show the evolution of the average magnetic moments, see Figs. S3(g,h), as function of the $U$ value. All of them increase in magnitude with increasing $U$.

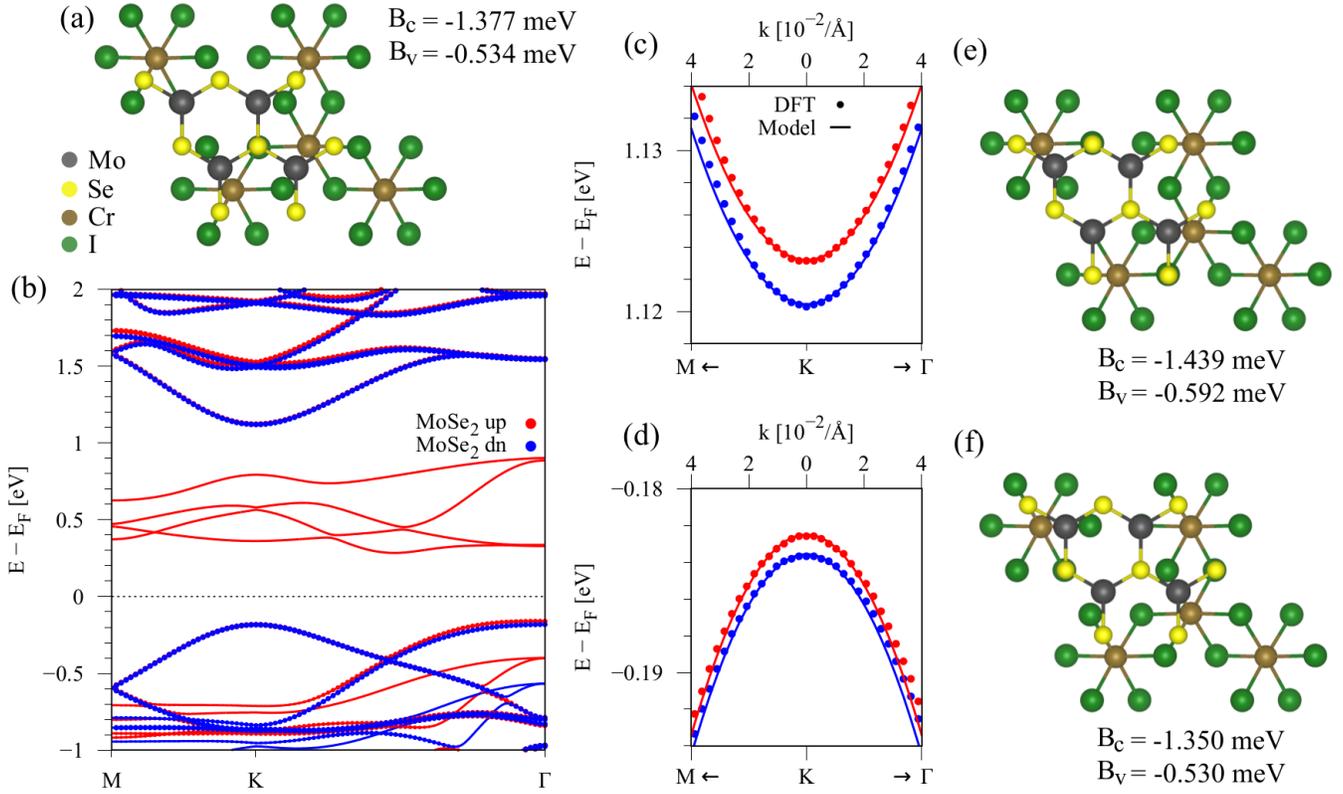

FIG. S2. (Color online) Calculated band structure without SOC and different geometries. (a) Different interface configuration of a 2 × 2 $MoSe_2$ cell above a 1 × 1 cell of $CrI_3$ and proximity exchange parameter values $B_c$ and $B_v$. (b-d) The band structure and zooms to the low energy bands with a fit to the model Hamiltonian, corresponding to the structure and parameters in (a). (e,f) The same as (a), but for other interface configurations.

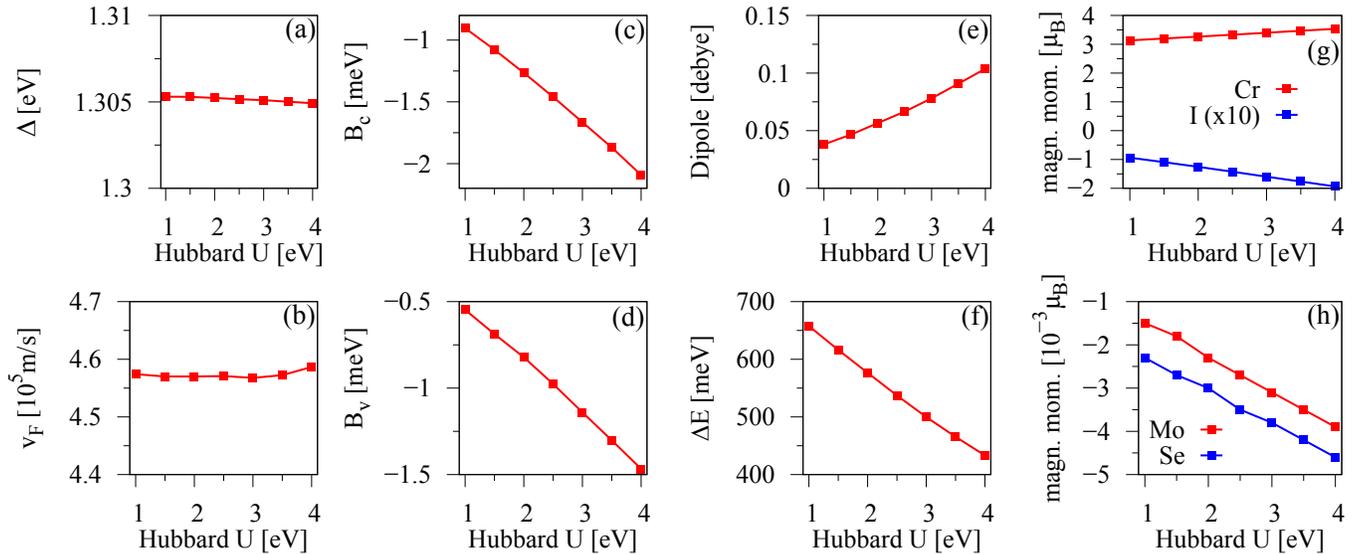

FIG. S3. (Color online) Fit parameters as a function of Hubbard $U$ parameter for $MoSe_2/CrI_3$ heterostructures for calculations without SOC. (a) The orbital gap $\Delta$ of the spectrum, (b) the Fermi velocity $v_F$, (c,d) the proximity exchange parameters $B_c$ and $B_v$, (e) the dipole of the heterostructure, and (f) the band gap $\Delta E$, as defined in the main text (see Fig. 1). (g,h) The average magnetic moments on the Cr, I, Mo, and Se atoms. We averaged over the Se atoms closer to the $CrI_3$ substrate. Magnetic moment of the I atom is enhanced by a factor of 10.

Even though, the proximity exchange parameters depend linearly on $U$, the order of magnitude does not change. Therefore we believe that our predictions of short-rangeness, gate, and twist effects will not be affected much; only the magnitudes may be a bit different.

### C. MoSe$_2$ on bilayer CrI$_3$

We consider MoSe$_2$ on bilayer CrI$_3$, when it is in the FM or AFM configuration, and fit the Hamiltonian to the low energy bands of the TMDC. The calculated band structures for MoSe$_2$/bilayer-CrI$_3$, including SOC, are shown in Fig. S4. The bands, originating from the TMDC, are barely different for the two magnetic configurations. The main difference, between the two magnetic configurations of the bilayer CrI$_3$ is, that the in-gap states from the CrI$_3$ are either fully spin-polarized or both spin species are present.

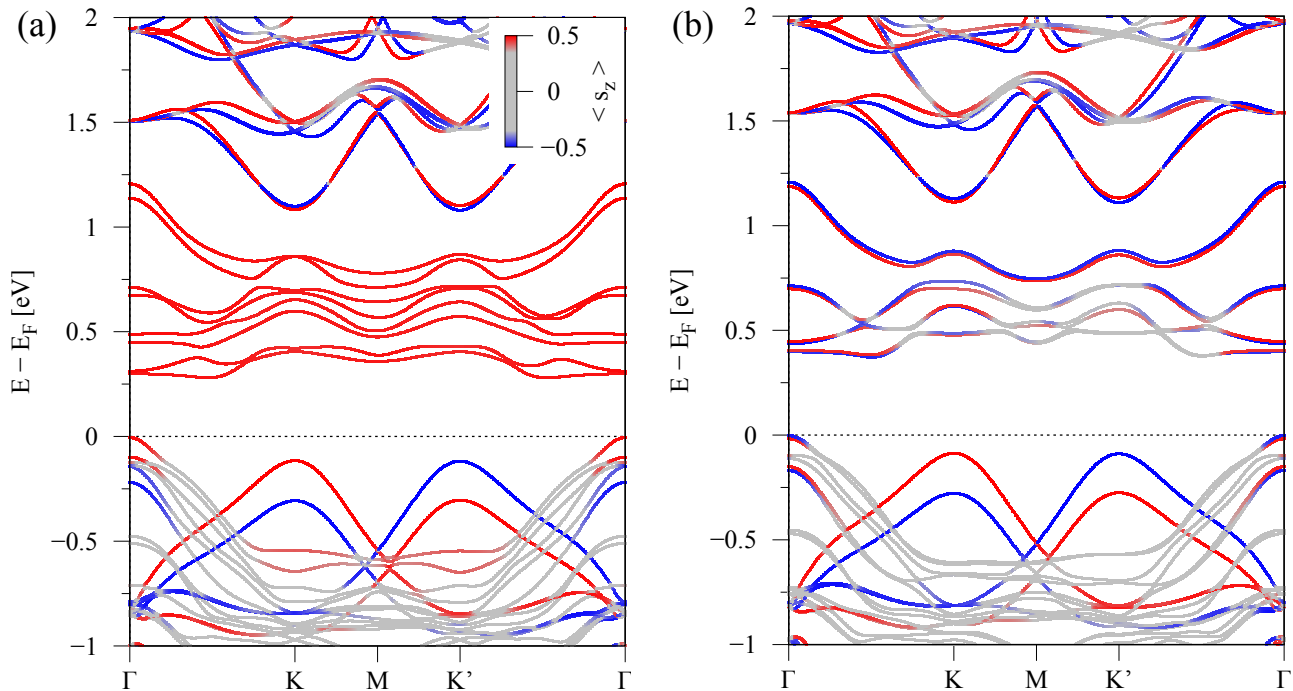

FIG. S4. (Color online) Calculated band structure of MoSe$_2$/bilayer-CrI$_3$ with SOC when the bilayer CrI$_3$ is in the (a) FM (↑↑) or (b) AFM (↑↓) configuration. Color corresponds to the $s_z$-expectation value. Zoom to the band edges are similar as shown for the monolayer CrI$_3$ case.

| configuration | FM (↑↑) | | AFM (↑↓) | |
|---|---|---|---|---|
| calculation | noSOC | SOC | noSOC | SOC |
| $\Delta$ [eV] | 1.307 | 1.303 | 1.307 | 1.303 |
| $v_F$ [$10^5$m/s] | 4.582 | 4.588 | 4.585 | 4.582 |
| $\lambda_c$ [meV] | - | -9.69 | - | -9.66 |
| $\lambda_v$ [meV] | - | 94.41 | - | 94.36 |
| $\lambda_R$ [meV] | - | - | - | - |
| $B_c$ [meV] | -1.766 | -1.542 | -1.703 | -1.408 |
| $B_v$ [meV] | -1.184 | -1.192 | -1.191 | -1.192 |

TABLE S2. Summary of the fit parameters for MoSe$_2$/bilayer-CrI$_3$ systems, where the bilayer-CrI$_3$ is in FM or AFM configuration, and arrows (↑,↓) denote the magnetization of the first and second CrI$_3$ layer below MoSe$_2$. The orbital gap $\Delta$ of the spectrum and the Fermi velocity $v_F$. The parameters $\lambda_c$ and $\lambda_v$ describe the SOC splittings, and $B_c$ and $B_v$ are the proximity exchange parameters for CB and VB. The dipole of the structure is 0.072 Debye for all configurations. The individual columns denote calculations without and with SOC.

The fit results are summarized in Tab. S2. A naive expectation would be that, depending on the total magnetization of the bilayer $CrI_3$, we can enhance or reduce proximity exchange in the TMDC, compared to the monolayer $CrI_3$ case. We can see that the FM ($\uparrow\uparrow$) coupled bilayer gives almost no difference in the fit parameters compared to the monolayer case. By switching the magnetization to AFM ($\uparrow\downarrow$), the magnitude of the exchange parameters stay similar. *Even though the AFM coupled bilayer $CrI_3$ has no net magnetization, we still find proximity exchange in the TMDC.* We conclude that only the FM $CrI_3$-layer closest to the TMDC is responsible for the proximity exchange, which is truly a short range effect, in agreement with previous thoughts[7].

### D. Bilayer MoSe$_2$ on CrI$_3$

In the main text, we have shown in Figs. 3(a-e) the calculated band structure and geometry of bilayer $MoSe_2$ on $CrI_3$. For further insight, we show in Fig. S5 the calculated spin-resolved density of states (DOS) for each individual layer. Especially interesting is the finite spin-up DOS around the Fermi level of the first $MoSe_2$ layer above $CrI_3$. Due to hybridization, the spin polarized in-gap states, resulting mainly from the $CrI_3$, have an additional small contribution from the first $MoSe_2$ layer. This result further confirms the short range proximity exchange coupling.

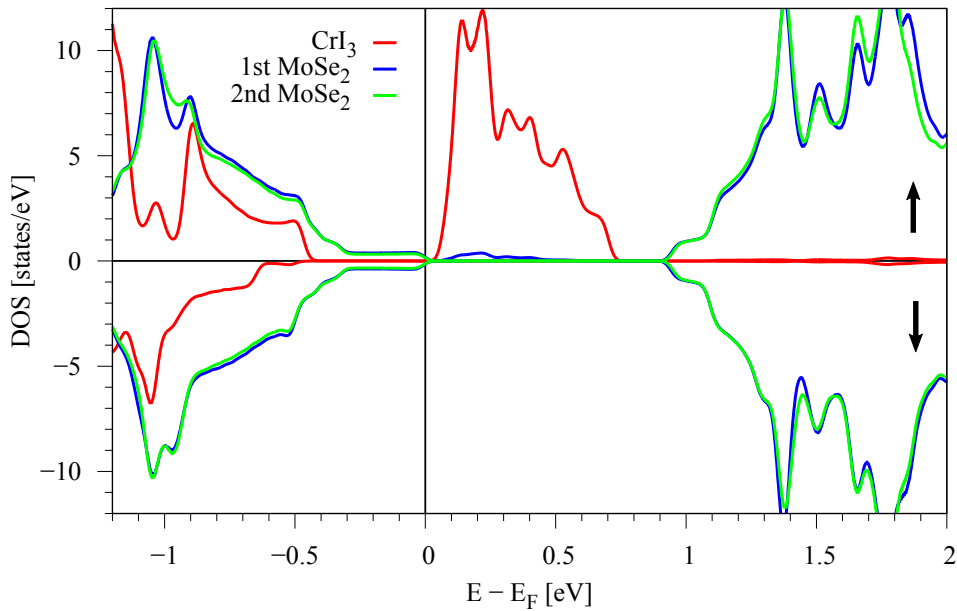

FIG. S5. (Color online) Calculated spin resolved density of states (DOS) for bilayer $MoSe_2$ on $CrI_3$. Different colors correspond to the total DOS of the different layers. Positive (negative) DOS values stand for spin up (down).

### E. hBN buffer layer

Considering the air stability of $CrI_3$, it may be crucial to use a hBN capping layer, to protect it from environmental impurities and degradation. Does a TMDC, on top of the hBN protected $CrI_3$, still experience proximity exchange? To answer this, we consider a stack out of a $4 \times 4$ cell of $MoSe_2$, a $5 \times 5$ cell of hBN and a $2 \times 2$ cell of $CrI_3$. We change the lattice constant of $MoSe_2$[5] to $a = 3.314$ Å, the one of hBN[42] to $a = 2.651$ Å, and the one of $CrI_3$[3] to $a = 6.627$ Å, resulting in strains of 0.79%, 5.87%, and -3.49%. Even though the strain of the hBN is quite large, we can give an insight on the effect of the buffer layer on the proximity exchange. The geometry of $MoSe_2/hBN/CrI_3$ contains 130 atoms in the supercell. Again, before calculating the band structure, we allow for relaxation of the whole geometry. Interlayer distances are relaxed to 3.46 Å (3.36 Å), between the hBN and $CrI_3$ ($MoSe_2$). Self consistent calculations are performed with a $k$-point sampling of $12 \times 12 \times 1$

In Fig. S6 we show the geometry and the calculated band structure of the $MoSe_2/hBN/CrI_3$ stack. Similar to the case without the hBN layer, the spin polarized bands of $CrI_3$ reside within the band gap of $MoSe_2$. The zooms around the band edges show, that the hBN buffer layer protects the TMDC from proximity exchange. The spin splittings

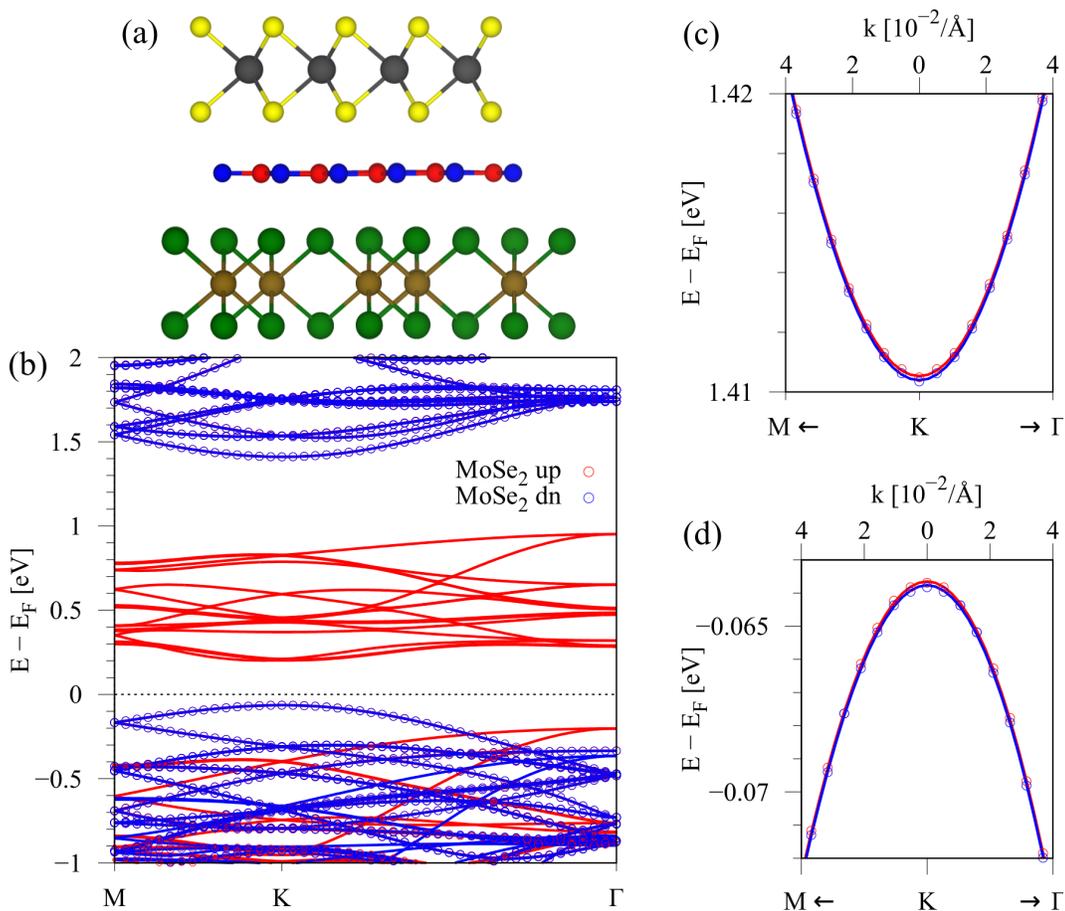

FIG. S6. (Color online) Calculated band structure without SOC and geometry. (a) Side view of the $MoSe_2/hBN/CrI_3$ stack. (b) Band structure along high symmetry lines. The bands corresponding to $MoSe_2$ are emphasized by red (spin up) and blue (spin down) spheres. (b) Zoom to the CB edge around the K point. (c) Same as (b), but for VB edge.

diminish to about 100 $\mu$eV, compared to several meV without the hBN layer. The fit parameters for this case are $\Delta = 1.474$ eV, $v_F = 4.638 \times 10^5$ m/s, $B_c = -0.069$ meV, $B_v = -0.057$ meV.

In Fig. S7 we show again the band structure of the $MoSe_2/hBN/CrI_3$ stack, now with focus on the hBN layer. The bands of hBN are located far away from the Fermi level, and are strongly hybridized with the bands of $CrI_3$ and $MoSe_2$, see Figs. S7(b,c). The band structure, projected on the states of hBN in this heterostructure geometry, is barely reminiscent of the monolayer hBN dispersion. However, the analysis of the hBN bands can be important for interpreting tunneling experiments in such a geometry. Especially the strong hybridization is an indication that electrons and holes can tunnel from the $CrI_3$ through hBN into the TMDC.

## S5. READING AND WRITING MAGNETIC STATES

In the following, we propose a geometry, that can optically or electrically readout the magnetic state of bilayer $CrI_3$. Suppose that a bilayer $CrI_3$ has two proximity coupled TMDCs on both sides, see Fig. S8 (a). The top (bottom) TMDC couples only to the top (bottom) $CrI_3$ of the bilayer. If the bilayer $CrI_3$ is FM coupled, both TMDCs experience the same proximity exchange, see Fig. S8 (a), which can be detected, e.g., optically via PL spectrum or electrically via characteristic magnetotransport phenomena. Especially when two different TMDCs are used, two distinct energy peaks can be seen in the PL spectrum, each of which is valley polarized, corresponding to the specific proximitization. By switching the magnetization of the top $CrI_3$ layer, see Fig. S8 (b), which is experimentally possible via optics, magnetism or gating[7,43–45], the proximity exchange induced in the top TMDC also switches sign, resulting in the opposite valley polarization. Therefore, all possible magnetic states (FM or AFM) of the bilayer $CrI_3$ can be distinguished by detecting proximity exchange in the TMDC layers.

Recently, photoinduced magnetization switching of the $CrI_3$ layer underneath a TMDC[7] was shown, by tuning the

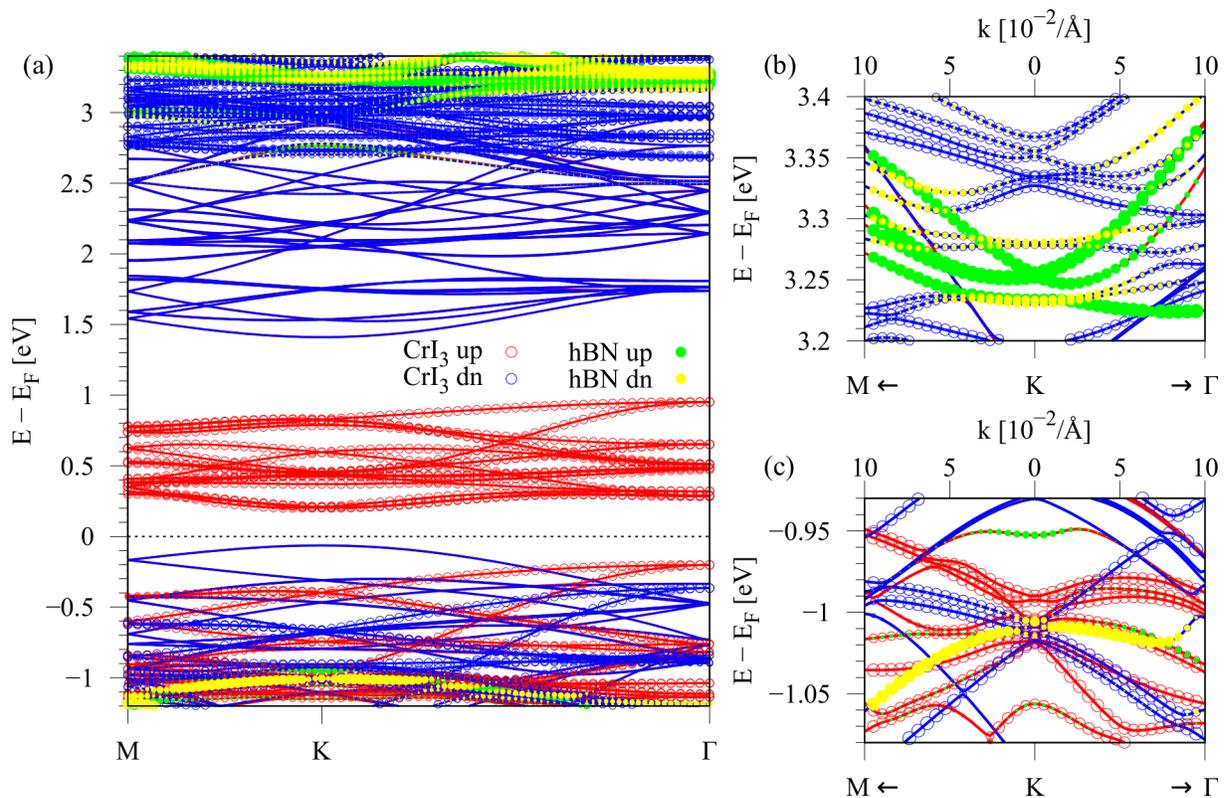

FIG. S7. (Color online) Calculated band structure without SOC of the MoSe$_2$/hBN/CrI$_3$ stack. (a) Band structure along high symmetry lines, where the bands corresponding to CrI$_3$ are emphasized by red (spin up) and blue (spin down) solid spheres. The bands corresponding to hBN are emphasized by green (spin up) and yellow (spin down) spheres. (b,c) Zoom to the band edges around the K point corresponding to hBN.

laser excitation power. The optical induced carrier density $n_o$ depends on the laser power, the excitation energy, the absorption coefficient and the area of the laser spot, as reported for semiconductor nanowires[46]. The electrical gate induced magnetization switching occurs for threshold densities of $n_t \approx 2 \times 10^{13} \text{cm}^{-2}$[45]. Obviously, magnetization switching in CrI$_3$ occurs, when the induced optical density satisfies $n_o \geq n_t$. However, as noted by Seyler et al.[7], it is unclear what exactly contributes to $n_o$. We believe that the switching appears due to a combination of direct optical excitation in CrI$_3$ and from an indirectly induced exciton contribution transferred from the TMDC, see Fig. S9.

Let us now speculate on a scheme, where magnetization switching happens solely due to the induced exciton density. If one resonantly excites the TMDC with circularly polarized light, addressing only one valley, one can switch the magnetization of the CrI$_3$ layer underneath. Thereby, the proximity exchange in the TMDC switches sign, the band gap in the corresponding valley changes, and the optical excitation in the TMDC is getting off resonance. Without the induced exciton density, the bilayer CrI$_3$ then relaxes back to its initial magnetic state. The TMDC is again in resonance, as the original band gap in the valley is restored, and the process starts anew, resulting in oscillatory magnetization switching. Especially materials with similar properties and a larger band gap than CrI$_3$ should be suitable to observe this effect, since the direct optical density is then decoupled from the indirect excitonic one.

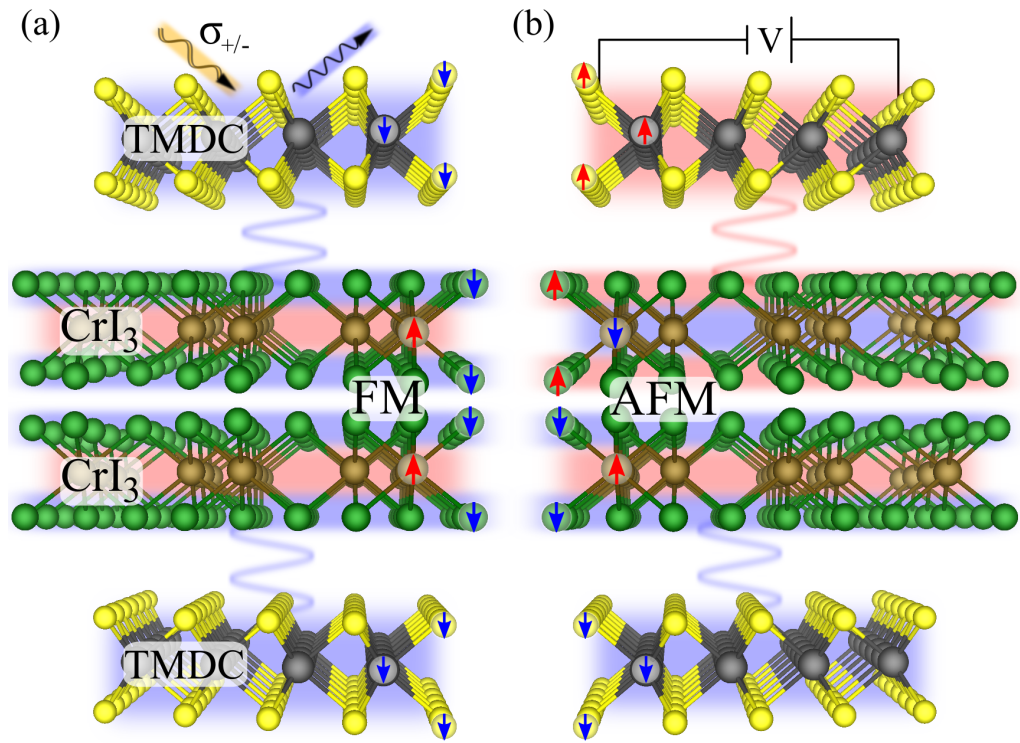

FIG. S8. (Color online) Optical and electrical readout of FM and AFM state. Sketch of the proximity exchange, originating only from the CrI$_3$ closest to the TMDC. The colors correspond to the magnetization direction of the atomic layers ($\uparrow$ = red, $\downarrow$ = blue). Top (bottom) TMDC layer is coupled only to the top (bottom) layer of bilayer CrI$_3$. In the FM state (a), proximity exchange for both TMDCs is similar, while switching the magnetization in one of the CrI$_3$ layers, leads to an AFM state (b) and the proximity exchange in the corresponding TMDC switches sign. Proximity exchange of the TMDC can be detected (a) optically by PL or (b) electrically in magnetotransport.

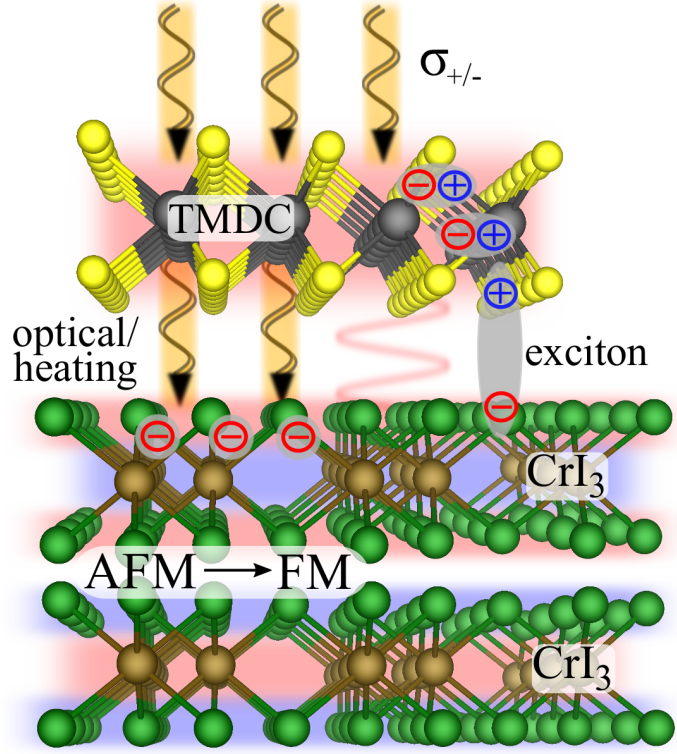

FIG. S9. (Color online) Shining light on the heterostructure directly creates (i) an exciton density in the TMDC and (ii) carriers in the CrI$_3$ layer. The combination of direct optical excitation in CrI$_3$ and from an indirectly induced exciton contribution transferred from the TMDC, switches the magnetization of the top CrI$_3$ layer and induces transitions from the AFM ground state to a FM state of the bilayer CrI$_3$.

\clearpage

\end{document}